\documentclass[journal]{new-aiaa}

\usepackage{nomencl}
\usepackage{ifthen}
\usepackage{bm}
\usepackage{diagbox}
\usepackage{booktabs}
\usepackage{caption}
\usepackage{subcaption}
\usepackage{cleveref}
\usepackage{color}
\usepackage{graphicx}
\usepackage[section]{placeins}
\usepackage{hyperref}
\usepackage{multirow}
\usepackage{rotating}
\usepackage{pdflscape}
\usepackage{lineno}
\usepackage{comment}

\usepackage{amsmath,amssymb,mathrsfs,amsfonts,dsfont}

\DeclareMathOperator*{\argmax}{arg\,max}
\DeclareMathOperator*{\argmin}{arg\,min}

\newcommand{\norm}[1]{\left\lVert#1\right\rVert}
\newcommand{\bchi}{\ensuremath{\bm{\chi}}}
\newcommand{\overbar}[1]{\mkern 1.2mu\overline{\mkern-1.2mu#1\mkern-1.2mu}\mkern
1.2mu}

\graphicspath{{./figures/}}

\begin{document}
\title{Improved Surrogate Modeling using Machine Learning for Industrial Civil Aircraft Aerodynamics}



\author{Romain Dupuis,\footnote{PhD. Student, Embedded Systems Department, 118 route de Narbonne, Toulouse.} Jean-Christophe Jouhaud,\footnote{Senior researcher, CFD Team - AAM Group, 42
Avenue Gaspard Coriolis, Toulouse.}}
\affil{IRT Saint Exup\'{e}ry, Toulouse, 31 432, France \\ CERFACS, Toulouse, 31 057, France}
\author{Pierre Sagaut\footnote{Professor and Director, Aix
Marseille Univ, CNRS, Centrale Marseille, M2P2 UMR 7340, 13451 Marseille cedex,
France.}}
\affil{Aix Marseille Univ, CNRS, Centrale Marseille, M2P2 UMR 7340, 13451
Marseille cedex, France}

\maketitle

\begin{abstract}
Predicting and simulating aerodynamic fields for civil aircraft over wide flight envelopes represent a real challenge mainly due to significant numerical costs and complex flows. Surrogate models and reduced-order models help to estimate aerodynamic fields from a few well-selected simulations. However, their accuracy dramatically decreases when different physical regimes are involved. Therefore, a method of local non-intrusive reduced-order models using machine learning, called Local Decomposition Method, has been developed to mitigate this issue. This paper introduces several enhancements to this method and presents a complex application to an industrial-like three-dimensional aircraft configuration over a full flight envelope. The enhancements of the method cover several aspects: choosing the best number of models, estimating apriori errors, improving the adaptive sampling for parallel issues, and better handling the borders between local models. The application is supported by an analysis of the model behavior, with a focus on the machine learning methods and the local properties. The model achieves strong levels of accuracy, in particular with two sub-models: one for the subsonic regime and one for the transonic regime. These results highlight that local models and machine learning represent very promising solutions to deal with surrogate models for aerodynamics.

\end{abstract}

\section{Introduction}

During the aerodynamic design of aircraft, an important amount of representative parameters of the flight missions must be taken into account. Computational Fluid Dynamics (CFD) is a powerful tool to support aircraft design. In particular, high-fidelity methods, such as Reynolds Averaged Navier Stokes (RANS) simulations, are usually able to reproduce complex aerodynamics. However, a huge number of parameter combinations is impossible to consider with the current high-fidelity tools given the dimension of the flight envelope and the computational cost associated with these tools. A classical solution consists of replacing the expensive high-fidelity model with a surrogate model. The latter is a mathematical approximation much faster to run built from a sampling of high-fidelity simulations. This substitution is achieved naturally at the expense of model precision and the main priority is the trade-off between the precision and the computational cost.

A particular kind of surrogate model is referred to as Reduced-Order Model (ROM). The central idea is to project all the degrees of freedom of the solution onto a set of basis functions of a smaller dimension. Reducing the number of degrees of freedom tackles computational
complexity while keeping interesting physical integrity. Different methods exist to compute the basis functions, such as the Krylov-subspace methods~\cite{Grimme1997} or the Proper Orthogonal Decomposition (POD)~\cite{Holmes1996}. We selected this latter in our approach. In the reduced space, the coefficients of the projection, also referred to reduced coordinates, have to be determined. Three main methods exist to compute these coefficients: projection-based methods using the governing equations, residual minimization and data-fitting models. Projection-based methods~\cite{Epureanu2003, Lieu2006} have the advantage to retain some of the physics from the governing equations and to give rigorous error bounds and error estimation~\cite{Veroy2005}. However, they require to have access to the governing equations inside the source code. This approach cannot be used for commercial softwares for which the source code is not available and becomes very difficult for complex and legacy codes. Residual minimization uses instead other equations, usually simpler. For instance, reduced coordinates from RANS simulations can be computed by minimizing the residual of the Euler equation~\cite{Zimmermann2010}. However, the definition of this residual expression plays a crucial role in the behavior of the system~\cite{Alonso2011, Amsallem2013}. Moreover, a part of the physical phenomena can be cut off by simplified governing equations. These severe drawbacks motivate the use of non-intrusive methods needing only data computed by the full-order model to approximate the reduced coordinates. Various techniques have been considered to infer the reduced coefficients: spline interpolation~\cite{BuiThanh2003parametricPOD}, Kriging~\cite{Fossati2015}, or radial basis functions~\cite{xiao2015non}. Interpolated ROMs are very popular non-intrusive methods, especially for steady and parametric aerodynamic problems~\cite{BuiThanh2003parametricPOD, Braconnier2011, Dolci2016, Malouin2013, Qiu2015}. That is the reason why we propose to use them in this paper.


Nevertheless, the prediction of solutions with bifurcations or a mixture of physical regimes can be the source of high discrepancies for non-intrusive ROMs, since the solution can change considerably as input parameter values change. This typical behavior is proved to be complex for surrogate models. Two explanations can be given to this problem. First, discontinuous features dominate the POD modes and thus small errors in the multivariate regression step amplify POD modes associated with a physical regime that does not exist for the considered prediction. The analysis performed by~\cite{Li2016} confirms these different insights. The authors highlight the poor capability of the interpolation ROM to properly predict the flow solution in extrapolation regions when high discontinuities are present. The evident mismatch between POD, a linear method, and the reconstruction of nonlinear responses is also emphasized. The second explanation involves a set of basis functions that is not representative enough due to a lack of relevant data. In addition to those comments, the mathematical properties of the POD and those of the equations can be compared. As underlined in~\cite{Noack2016}, the POD exhibits elliptic properties while aerodynamics is characterized by a hyperbolic part in the equations. This mismatch is the root cause of most of its weaknesses. In that respect, an additional computationally intensive endeavor is required to properly train a model built on a training set mixing subsonic and transonic flows. This ensures that the linear combination of the reduced basis functions is able to reproduce any shapes and that the regression model is well trained. Otherwise, the ROM fails to produce accurate responses with a restricted number of dominant modes for predictions in highly nonlinear regions and not directly in the neighborhood of the snapshots~\cite{Malouin2013, Qiu2015, Dolci2016}. However, this need for important computational resources goes against the principle of the parametric ROM that aims at building surrogate models with a favorable trade-off between computational cost and accuracy.

For this reason, specific local surrogate models dealing with various physical regimes have been developed. The main idea is to decompose the spatial grid or the parameter space in subsets. As regards the spatial separation,~\citet{Lucia2003} demonstrate the ability of methods using two different models on two-dimensional flows with moving shock waves, where a full-order model is used within the region of the flow containing a moving shock. But the main issue of such  strategy is to determine for each mesh node which model to use.~\citet{Constantine2012} identify regions with high parameter gradients. For each
point of the domain, the gradient is computed and if it exceeds a given threshold, the point considered of poor quality is tagged as requiring a specific correction. Shocks are well reconstructed with this strategy. However, the method is only applied on a one-dimensional parameterized hyperbolic conservation law.~\citet{Bergmann2018} split the computational domain with a leave-one-out strategy. The a priori estimated error is computed on the whole domain. If the value exceeds a given threshold, the point is associated with the high-fidelity region managed by the full-order model. Obviously, the value given to the threshold plays a crucial role, as a too small value leads to a high fidelity region covering the whole domain. It can be noted that an overlapping zone is also introduced, defined by a second threshold value performing an optimal transition between full-order models and ROMs. This method has been used to solve aerodynamic shape optimization of a car section and shows a reduction $5\%$ of the drag coefficient.~\citet{Iuliano2013} conceive a zonal method in order to solve optimization problems. In that case, the high fidelity region is manually defined near the wall. The surrogate-based optimization coupled with the zonal CFD/POD model performs very well on the shape optimization of a RAE2822 airfoil, especially by finding a profile with a weak shock.

The other kind of decomposition based on the parameter space aims at specializing local ROMs for a single kind of pattern or physical regime. Combining these different local models improves the overall accuracy, as local models are more specialized and trustworthy but only on smaller parts of the parameter space. In the context of scalar models, this type of approach is called mixture of experts and has been introduced in the 1990s to solve regression and classification problems~\cite{Jordan1994, Yuksel2012}. Mixtures of expert model are competitive for regression problems with non-stationary and piecewise continuous data. That is why it appears well suited to applications featuring discontinuities.
For example,~\citet{Bettebghor2011} have developed a specific strategy, based on a Gaussian mixture of experts, able to approximate discontinuous functions. It turns out that mixtures of experts dramatically improve the accuracy of the predictions compared to the best global surrogate model. Recently, mixtures of experts have also been used to evaluation aircraft fuel burn over a complete mission~\cite{Liem2015}. The authors highlight the clear superiority of mixtures of experts compared to conventional surrogate models to approximate aerodynamic coefficients with complex shapes. Both examples presented here rely extensively on machine learning methods, in particular unsupervised learning for the separation of the solutions and supervised learning, in the case of~\citet{Liem2015}, in order to associate new predictions with the correct cluster.

The parameter space decomposition has recently emerged in ROM literature~\cite{Washabaugh2012, Amsallem2012, Zhan2015}. The central idea is the same as for scalar problems: considering only a restriction of the full set of training snapshots in order to have local surrogates more adapted to the solution. The method can take different forms depending on the type of problem or surrogate model. One may argue that the capability of compression of POD is weakened by increasing the number of bases and that the robustness of the method computing the reduced coordinates can decrease with the reduction of training samples, which are spread on different models. But on the contrary, for cases with various physical regimes, the local models enable a clear separation of the phenomena, which improves the surrogate models. The advantage is two-fold: the local POD basis is more representative and interpolation tools (or projected equations) gain in flexibility by being heterogeneous on the parameter space. The parameter space decomposition can be seen as a blessing of abstraction, countering the curse of dimensionality~\cite{Bellman1961}: if the local models are more adapted to physics, they need to learn fewer structures, compared to global approaches. Moreover, as mentioned previously, the POD shows elliptic properties, thus these kinds of local approaches can demonstrate similarities with the discontinuous Galerkin method to tackle hyperbolic problems. As regards intrusive ROMs, the first attempts of parameter domain decomposition do not use machine learning. \citet{Cizmas2008} propose a method in order to accelerate ROMs. The transient part is separated from the steady part. Two splitting methods are tested: they monitor the variation of the reduced coordinates and the ratio between CPU time and physical time. These two strategies identify successfully the transition between the two regimes by using a given threshold defined in advance.~\citet{Haasdonk2011} introduce a recursive subdivision of the parameter domain based on quadtree. For each leaf element of the grid, if the error bound or the size of the basis exceed given thresholds, the element is refined into several subdomains. Then, the process is restarted until all the thresholds are not exceeded. However, such a method based on quadtree is very sensitive to the dimension of the parameter space. 

Strategies relying on machine learning have been developed in order to build intrusive or non-intrusive ROMs: unsupervised learning algorithm groups similar snapshots into clusters, local ROMs are built on each cluster, and a classifier assigns the input parameter of each prediction to the best suited local ROM. This strategy has already been applied successfully to build intrusive parametric ROMs for fluid-structure-electric interaction problems~\cite{Amsallem2012} and aerodynamic problems~\cite{Washabaugh2012}.~\citet{Peherstorfer2014} propose a sophisticated method by coupling parameter domain decomposition and discrete empirical interpolation method. It widely reduces the prediction error of the ROMs in the case of a simple reactive flow. Finally, to the best of the author's knowledge, the only method of non-intrusive local ROMs has been developed in the context of the aero-icing certification~\cite{Zhan2015, Zhan2016}.

The Local Decomposition Method (LDM), used in this paper, extends the
classical POD/interpolation ROM by splitting the parameter space into subsets. This approach improves the prediction capability of surrogate models for quantities of interest with very different behaviors within the input parameter space. LDM has already been presented in previous papers and applied successfully to two and three dimensional external aerodynamic cases~\cite{Dupuis2018, Dupuis2018b}. The local POD bases are determined using machine learning tools yielding to more flexible behaviors bringing out a precise delimitation of the physical regimes. Clustering methods aim at separating the snapshots in function of the different flow shapes. A specific shock sensor helps to achieve proper separations of the physical regimes. Indeed, LDM has mainly been developed to deal with external aerodynamics with flow separations and a mix between the subsonic and transonic regimes. This specific shock sensor measures the nonlinearities and sharp gradients of the flowfield. Thus, the different regimes are no more mixed in the POD basis. As regard the prediction of untried
sets of parameters, the parameter space is divided into several domains by a classifier according to the clustering of the snapshots. Therefore each
region of the parameter space is associated with a local reduced-order model and its
respective subspace, allowing to classify the input parameter space to the
right physical regime. Finally, resampling is carried out by identifying the subspaces with the highest entropy. Extra snapshots are added in these specific subspaces with the objective to minimize the redundancy of the sampling, thus increasing the accuracy of the
surrogate model for a minimum computational cost.

In addition to the specific treatment of complex flow regimes, it appears that industrial-like configurations are rarely investigated in the literature. Only few examples of complex applications can be cited.~\citet{Fossati2015} applies non-intrusive ROMs to various aerodynamic cases of industrial relevance for viscous turbulent flows: a truncated wing, an aircraft configuration, and a helicopter rotor in hover. In addition to the encouraging results in term of ROM accuracy, a complete framework of ROM has been implemented and employed to solve a aero-icing problem on a configuration with fuselage, wing, pylon, and engine nacelle, highlighting the capability of such method to explore both aerodynamic and icing envelope~\cite{Fossati2013}.~\citet{Roy2018} perform an uncertainty quantification analysis on a LS89 blade cascade. Even if the geometry remains simple, the authors use Large Eddy Simulation on a $20$ million cell mesh. Such computational costs represent a real challenge but they are necessary to demonstrate capabilities of the non-intrusive ROMs to deal with complex cases, in particular for uncertainty quantification. In the context of the Digital-X project, various ROM methodologies have been applied to a configuration of XRF-1 aircraft, a 3D test case of a transonic wing-body transport aircraft~\cite{dlr2016}. For example, ROMs are integrated into a framework to compute aeroelastic loads at different flight conditions. The structural displacements are in good agreement between coupled high-fidelity models and ROMs, even if the prediction of the pressure distribution shows significant discrepancies on the suction surface, near the shock wave region. Regarding supersonic flows,~\citet{Mifsud2010} couple POD and radial basis functions for a parametric study of weapon aerodynamics. They consider a three-dimensional flow around a fin-stabilized missile with three parameters: Mach number, incidence, and flare base radius. The results suggest that a reliable low-cost high-fidelity tool can be built: from a small number of simulations, the pressure and density contours are well predicted for a specific flow regime.~\citet{Chen2015} propose an aerothermodynamic application of interpolation-based ROM for hypersonic vehicles. The geometry is based on the Lockheed F-104 Starfighter wing. Thermal and chemical non-equilibrium models are considered in the FOM. The mesh counts $819.000$ nodes and $3$ parameters vary: Mach number, angle of attack, and altitude. ROMs are validated by leave-one-out cross validation. Predicting the temperature requires an important number of samples, between $100$ and $300$, in order to reach satisfactory accuracy whatever methods used. Margheri and Sagaut~\cite{Margheri2015, Margheri2016} combine POD, Kriging, and sensitivity decomposition with the objective of quantifying the uncertainties of urban pollutant dispersion using Lattice Boltzmann Methods. The more challenging computation simulates a full urban area from two to five input variables. The computational grids count $6$ millions of cells and a single simulation requires about $1.000$ CPU hours. In the present paper, we propose to apply the LDM method to complex and industrial relevant case: the XRF-1 aircraft configuration. A very fine mesh is investigated with more than $120$ million nodes and $7$ different input parameters including engine conditions in order to be close to real industrial process.

This paper aims to present improvements to the LDM and its application to a industrial-like aerodynamic application. It is organized as follows:~\autoref{sec:ldm} gives a quick overview of the LDM approach. Then, \autoref{sec:improvement} presents various improvements to the method, in particular how to handle the interface between different local ROMs and how to control the resampling. Then, results from the three-dimensional XRF-1 aircraft configuration are presented in~\autoref{sec:XRF1}, illustrating the capability of the LDM to deal with both industrial constraints and complex physical regimes. Finally,~\autoref{sec:conclusion} provides a summary and the conclusions.
\section{Local Decomposition Method: Local ROMs using Machine Learning}
\label{sec:ldm}
The Local Decomposition Method (LDM) presented in this section aims at improving the predictions of surrogate models for quantities of interest with very different behaviors within the input parameter space. Instead of a unique global POD basis, several local bases are determined using machine learning tools yielding to more flexible behaviors bringing
out a precise delimitation of the physical regimes. This method has mainly been developed in order to deal with external aerodynamics with flow separations, a mix between the subsonic and transonic regimes. However, it could be envisioned to generalize the LDM to other problems characterized by bifurcating solutions. By this way, the LDM is expected to reduce the computation budget needed to build the training samples. In addition, separating the snapshots into several cluster can help to improve the physical understanding from numerical simulations. LDM presented in~\cite{Dupuis2018, Dupuis2018b} is extended in the present work by improvements of the algorithm (in the control of the clusters, the parallel distribution of the computations, and the management of the interfaces) and by an industrial application. The general method is only briefly summarized as details can be found in the previous papers.

Notations are first introduced. The goal of ROMs is to substitute the CFD model in order to estimate a field of specific quantity of interest $f$ by a numerical approximation $\widetilde{f}$. The quantity of interest, for example a pressure field, is written as a function of two inputs: the mesh coordinate $\bm{x}\in \mathds{R}^{3}$ and a
vector of the input parameters $\bchi \in \mathds{R}^{p}$ where $p$ is the number of input parameters. The function $f$ is evaluated on a mesh $\Omega$ of size $d$ formed by a list of three-dimensional coordinates, such as $\Omega = [ \bm{x_1}, \cdots, \bm{x_d}
]^T \in \mathds{R}^{d\times 3}$, where $\bm{x_i}$ indicates the coordinate of
the $i$-th point of the mesh. The components of the vector $\bm{\chi}$
correspond to specific attributes of the CFD simulations, such as the value of a
boundary condition, the nature of the flow or a numerical parameter.
Building a surrogate model is based on training samples (also called training
set, training data or training sampling), generated from results of
several CFD simulations. In other words, quantities of interest
are computed on the mesh for $n$ different training input parameters
$\bchi_{\bm{t}}$, forming the training set of input parameters $ \bm{X_t} = [
\bchi_{\bm{t_1}}, \cdots, \bchi_{\bm{t_n}} ]^T \in \mathds{R}^{n \times p}$ where $\bchi_{\bm{t_i}}$ is the $i$-th sample of the training set. In the interests of simplifying notations, the mesh is omitted as it remains constant in our applications. The vector-valued function $\bm{f}$ for the training set is defined by $\bm{f(\bchi_{\bm{t}})} = \left[ f(\bm{x}_{\bm{1}}, \bchi_{\bm{t}}), \cdots , f(\bm{x}_{\bm{d}}, \bchi_{\bm{t}}) \right] \in \mathds{R}^d $. The training snapshot matrix $\bm{S_t} \in \mathds{R}^{n \times d}$ regroups the quantities of interest on the mesh and each row is defined by $\bm{s_{ti}} =  \bm{f}(\bchi_{\bm{t_i}}) \in \mathds{R}^{d} $. Once trained, the surrogate model aims at computing the snapshot matrix of the predictions $\bm{S_p}$ for the set of prediction sample $\bm{X_p} = [ \bchi_{\bm{p_1}}, \cdots, \bchi_{\bm{p_m}} ]^T$, where $m$ is the number of predictions and $\bm{\widetilde{f}}$ is the vector-valued function of the
surrogate model. Usually, the fluctuating quantity $\bm{f'}$ is analyzed instead of the direct function $\bm{f}$, where the mean snapshot has been subtracted $\bm{f}'(\bchi_{\bm{t}}) = \bm{f}(\bchi_{\bm{t}}) - E[\bm{f}(\bchi_{\bm{t}})]$ from $\bm{f}$. 
\subsection{POD-based reduced order models}

\subsubsection{Proper Orthogonal Decomposition}
POD is a linear method based on the identification and extraction of
coherent structures by computing an optimum basis $\bm{\phi} = [\bm{\phi_1}, \cdots, \bm{\phi_K}]$ in term of energy representation~\cite{Cordier2007}. Thus, POD contains more energy than any other basis for a given number of basis vectors, called modes. The reduced-order process of POD results in the computation and the selection of a few number $K$ of these modes containing a very large amount of energy. The remaining modes are cut off. Finally, each basis vector $\bm{\phi_k}$ is associated with a scalar value $a_k$, also called reduced coordinate, which depends on the parameter space. The function of the quantities of interest can be expressed as:
\begin{equation}
 \bm{f}(\bm{\chi}) \approx \sum_{k=1}^{K}
 a_{k}(\bm{\chi}) \; \bm{\phi_k}.
 \label{eq:pod_fundamental}
\end{equation}

A possible way of expressing POD aims at maximizing the mean projection of the function of interest into the orthonormal basis function $\bm{\phi}$, leading to the following maximization problem~\cite{Fahl2000}:
\begin{equation}
\begin{aligned}
\max\limits_{\phi_1, \cdots,\text{ } \phi_K } \sum_{k=1}^K \left\langle|(\bm{S_i'},\bm{\phi_k})|^2 \right\rangle,
\\
\text{subject to } (\bm{\phi_i},\bm{\phi_j}) =  \delta_{i,j} \ \forall i,j \in [1,K]^2.
\end{aligned}
 \label{eq:pod_max_pb}
\end{equation}
where $\delta_{i,j}$ is the the Kronecker symbol satisfying $\delta_{i,j} = 1$ for $i=j$ and $\delta_{i,j} = 0$ otherwise. The maximization problem in ~\autoref{eq:pod_max_pb} gives rise to an eigenvalue problem~\cite{Henri2005, Volkwein2013}. The latter can be solved by the so-called snapshot method proposed by~\citet{Sirovich1986}. In practice, a Singular Value Decomposition (SVD) is used to compute the eigen matrix. An iterative approach exists to reduce the computational cost when adding extra snapshots~\cite{Melenchon2005}. The number $K$ of retained POD modes is determined by a heuristic criterion based on the energy ratio contained in the eigen values~\cite{Cordier2007, Sirovich1986}. Then, the reduced coordinates are computed using the orthonormal property of the POD basis. 


\subsubsection{Interpolation method by Gaussian Process}
Gaussian Processes (GP) belong to supervised learning
algorithms. They have been designed to solve two distinct types of problem: regression (Gaussian Process Regression) and classification (Gaussian Process
Classification)~\cite{Rasmussen2005}. The definition of classification and the use
of GP in this context will be discussed latter. This section focuses on
the Gaussian Process Regression (GPR). The latter is very similar to another method called Kriging, introduced in the 1970's~\cite{Krige}, and which becomes very popular in design and
 analysis of computer experiments~\cite{Sacks1989}, in particular to create surrogate models of costly simulations~\cite{Forrester2008}. GPs shows great capability to deal with nonlinear problems as long as the number of input parameters is small
(less than about 50)~\cite{Fang2005} and they also provide an estimation of
the modeling error.
 
 The final function $\bm{f}$ is fully approximated by building $M$ data-fit anisotropic Gaussian processes $\widetilde{a}$, one for each reduced coordinate, such as:
\begin{equation} 
 \bm{f}(\bchi)\simeq \overbar{\bm{f}} + \sum_{k=1}^{M}
\widetilde{a_k}(\bchi) \ \bm{\phi}_{k}, \, \forall \bchi \in \mathds{R}^p.
 \label{eq:pod_interpolation}
\end{equation}
As the behavior of reduced coordinate is unknown, a classical Radial Basis Function (RBF) kernel is chosen to represent the covariance function of the GP. This popular kernel provides several advantages: interesting mathematical properties (smoothness, infinite differentiability and analytical derivability), few hyper parameters to estimate, easily interpretable, and a universal kernel as any continuous function can be represented by the RBF kernel under specific conditions~\cite{Duvenaud2014}. A Maximum Likelihood Estimation (MLE)~\cite{Rasmussen2005} method has been selected to estimate the hyperparameter values in order to take advantage of the derivative information provided by the RBF kernel. The Limited-memory Broyden-Fletcher-Goldfarb-Shanno Bounded (L-BFGS-B) algorithm~\cite{Byrd-1995} is used to solve the optimization problem associated with the MLE method. The different GP models have been built thanks to the machine learning library scikit-learn~\cite{scikit-learn}.

\subsubsection{Design of Experiments to sample the simulations}
Determining input parameters of the set of experiments is performed by Design of Experiments (DOE) methods. Their application to numerical simulations has been generalized by~\citet{Sacks1989}. As the nature of the model is unknown, the design points should be evenly distributed throughout the parameter space in order to observe the response of the model at various conditions. Moreover, the samples must not collapse and each dimension must be explored as far as possible over the range of its variation, in particular two samples should not share a common value for a given dimension. The DOE aims at fitting a good trade-off between these two properties of space-filling and non-collapsing. In our method, we select low discrepancy sequences and in particular the Halton sequence for the DOE~\cite{Kalos2008}. It belongs to Quasi-Monte Carlo methods, shows good space-filling properties, and above all it has the interesting feature to be deterministic and sequential. It means the size of the sampling can be extended on purpose while keeping all the geometrical properties of the design. This property is of central interest to develop active learning and resampling strategies. Usually, stochastic methods and optimized designs does not provide iterative designs as the number of points is determined before the DOE generation, otherwise some properties of the design can be lost.

\subsection{Separation of the parameter space}
The parameter space must be split according to the nature of the flow in order to build specialized local POD-ROM. Three different tools are at the heart of the separation of the parameter space: i) a shock-sensor highlighting the difference between snapshots with different same physical behavior, ii) a clustering method identifying groups of similar snapshots, and iii) a classification algorithm allowing to associate future predictions with the proper physical regimes and clusters.

\subsubsection{Preprocessing step: the shock sensor}
The shock sensor is a mathematical transformation converting quantities of interest into a sensor of specific physical regimes. It has been first used to support numerical scheme and post-processing treatment highlighting shocks. The main goal here is to sharply differentiate physical regimes to ease the clustering of snapshots. Somewhat similar methods can be found in~\citet{Lorente2008} or~\citet{Constantine2012}, where the gradient is used to identify parts of the signal which can be improved by a specific correction.

External aerodynamics with subsonic and transonic regimes represents the target application in this paper. These flows are characterized by shock waves, that is why the Jameson's Shock
sensor~\cite{Jameson1981} is chosen as it detects large sudden changes in the quantity of interest. It is related to the second order derivative of the pressure:
\begin{equation}
\nu_i = \frac{| p_{i-1} - 2p_i + p_{i+1}|}{\epsilon_0 + |p_{i-1}|
+2|p_i|+|p_{i+1}|}, \ \text{with $i$ such that } \bm{x_i} \in \Omega \backslash \partial \Omega,
\end{equation}
where $\partial \Omega$ is the border of the mesh and $\epsilon_0$ a constant avoiding
division by $0$. One can note that other shock sensors could be used~\cite{Oliveira2009}.

\subsubsection{Clustering method}
Once a filter has been applied on data, structures must be discovered in order to separate the snapshots in function of the different flow shapes. The clustering identifies inherent groups in the input data based on similarity measures. It has been applied in various fields of application, such as image segmentation or data compression~\cite{Bishop2006}. Three fundamental clustering methods are usually considered for the LDM: K-means, Gaussian Mixture Model (GMM), and Density-Based Spatial Clustering of Applications with Noise (DBSCAN). Interesting readers can refer to~\citet{Bishop2006} and~\citet{Hastie2001} for a more in-depth presentation of other methods. The K-means algorithm has been selected for this study as one of the main goal was to assess the behavior of the method for different numbers of clusters. 

Clustering in high-dimension represents a very challenging problem related to the curse of dimensionality~\cite{Bellman1961}. Many different transformations have been developed  to find embedded non-linear manifolds within high-dimensional space, such as Locally linear embedding or Isomap methods~\cite{Bishop2006}. However, a classical POD has been applied on sensor data, mainly due to its simplicity and robustness compared to more complex nonlinear dimensionality reduction methods. The percentage of variance is set to $99\%$ and resulting subspaces from this transformation are grouped in matrix $\bm{B}$ such as:
\begin{equation}
\bm{B} = [\bm{b}_1 \ \cdots \ \bm{b}_n]^T, \ \bm{b_i} \in \mathds{R}^{M'},
\end{equation}
where $M'$ is the dimension of the cut-off in the PCA associated with $99\%$ variance.

The K-means algorithm is a very popular, simple, and easy to implement method. If the data $\bm{B}$ are splitted into $K$ clusters, each cluster $k$ is defined by a center $\bm{\mu_k} \in \mathds{R}^{M'}$ where $k \in \{1, \cdots , K\}$. The squared Euclidean distance is chosen as a similarity metric. The goal of the K-means method is to found the clusters minimizing the distortion measure $J$ defined by the squares of the distance of each point to the closest cluster center $\bm{\mu_k}$:
\begin{equation}
J= \sum\limits_{i=1}^N \sum\limits_{k=1}^{K} r_{ik} \norm{\bm{b_i} - \bm{\mu_k}}^2,
\label{eq:kmeans_def}
\end{equation}
where $r_{ik} \in \{0,1\}$ is the binary indicator variable. It is equal to $1$ if $\bm{b_i}$ belongs to the cluster $k$ and is equal to $0$ otherwise. The function $J$ can also be seen as an inertia where $\bm{\mu_k}$ corresponds to the center of mass of the cluster $k$. A two-step iterative process is used to optimize the cost function $J$:
\begin{itemize}
\item The \textbf{assignment step} determines the new values given to the binary indicator variables $r_{ik}$. If the centers are considered as fixed, the distortion measure $J$ is a linear function of $r_{ik}$. Consequently, terms of the sum can be minimized independently for each point in the data set, such as the new $r_{ik}$ is chosen as:
\begin{equation}
  r_{ik}=\begin{cases}
    1, & \text{if } k = \argmin\limits_{k' \in \{1, \cdots, K\}} \norm{\bm{b_i} - \bm{\mu_{k'}}}^2\\
    0, & \text{otherwise}
  \end{cases}.
\end{equation}
The expression can be interpreted easily: each point is assigned to its closest centroid.

\item The \textbf{update step} minimizes the value of $J$ which is a quadratic function of $\bm{\mu_k}$ with the $r_{ik}$ held fixed. If the derivative of $J$ is set to zero, the new centroids are expressed as:
\begin{equation}
\bm{\mu_k} = \frac{ \sum\limits_{k=1}^{K} r_{ik} \bm{b_i} } { \sum\limits_{k=1}^{K} r_{ik} }. 
\end{equation}
Once again, an interpretation can be given to the above expression: the new centroid $\bm{\mu_k}$ is simply the mean of all the data assigned to the cluster $k$.
\end{itemize}

The two-step process is stopped when the difference between old and new centroids is below a given threshold. Even if the K-means algorithm always converges~\cite{Bishop2006}, the function $J$ is generally non-convex and minimums may be local. For this reason, the K-means algorithm needs to be run with different initializations and the best local minimum is selected. At the end, the training snapshots are divided in K clusters, such as:
\begin{equation}
\{ (\bchi_{t_1}, c_{t_1}), \ \ldots \ , (\bchi_{t_n}, c_{t_n}) \}  \text{ with
} c_{t_i} \in [1, \cdots, K] , \ \forall i \in [1, \ n],
\end{equation}
where $c_{t_i}$ is the value of the cluster.

\subsubsection{A-priori association to each cluster via classification}
Once the simulations are clustered, the parameter space needs to be divided into several domains according to the clustering. A classifier assigns the input parameter of each new prediction to the best suited local ROM and to the right physical regime. The classifier aims at finding the different domains in the parameter space. A Gaussian Process classifier (GPC) is selected~\cite{Rasmussen2005} as they offer several advantages, such as: learning the kernel without cross validation, a fully probabilistic predictions, an integrated feature selection, and good performance when coupled with anisotropic kernels. Indeed, other algorithms such as the k-nearest neighbors method is very sensitive to the choice of k~\cite{Hastie2001} and the optimal value can change as the number of training samples may increase during the training, for instance with resampling. Thus this approach is not adapted to ROMs.

\subsection{Adaptive sampling based on the entropy}
The central idea of the sampling strategy for the LDM is to take advantage of the input space separation into several subsets. On the one hand, if the local ROM shows very complex shapes, extra snapshots must be added in order to reduced the interpolation error and to improve the POD basis. On the other hand, redundant information for smooth and linear behaviors must be avoided.

For these reasons, a criterion based on the compressibility of the information is used to identify the priority clusters. The global entropy $H$ measures the redundancy of information and can be computed on each cluster using the POD eigenvalues~\cite{Cordier2007}:
\begin{equation}
H = - \frac{1}{\log (n)}\sum_{i=1}^{n}p_i \log(p_i) \text{ with
} p_i = \frac{\lambda_i}{\sum\limits_{j=1}^{n}\lambda_j}.
\end{equation}
The adaptive resampling of the LDM assumes that the entropy and the structures of the system with discontinuities or high gradients are directly correlated. Thus, the probability to find new modes with a non-negligible amount of energy is expected to be greater for clusters with high values of entropy. Therefore, the samples from the resampling step are distributed in clusters with high entropy by considering the relative ratio of entropy of each cluster. 

\subsection{Recombination in a global model}
The final recombination step consists in assembling the local reduced-order
models in a single composite global model. It represents a challenging and crucial problem, as the classification near the decision boundary might be prone to errors due to the classifier. In addition reduced coordinates can be extrapolated in this region. Thus, a special attention is paid to the interface between clusters. Several solutions may be envisaged.

The simplest solution lets the hard or soft clustering managing the interface. Starting from the $K$ clusters, a
simple weighted sum is calculated using a 'hard' split:
\begin{equation}
 \widetilde{\bm{f}}(\bm{\chi}_{p_j}) = \sum_{k=1}^{K}
 \mathds{1}_{C_k}(\bm{\chi}_{p_j}) \widetilde{\bm{f}_k}(\bm{\chi}_{p_j}),
  \ \forall j \in [1, \cdots,m],
 \label{eq:ldm_final}
\end{equation}
where
\begin{equation}
 \mathds{1}_{C_k}(\bm{\chi})=\begin{cases} 
               1\text{ if } k = \argmax\limits_{i \in [1, K]} P(\bm{\chi}
               \in C_i)
               \\
               0 \text{ else }
            \end{cases},
\end{equation}
and $\widetilde{\bm{f}_k}$ refers to the local POD/GPR model built on the
$k$-th cluster. This sum provides a  prediction of all the input space, leading to a global model which is not differentiable at the interface. The differentiable
predictions require to use differentiable weighting functions, which is not the
case for $\mathds{1}$. The soft clustering~\cite{Bettebghor2011} substitutes $\mathds{1}_{C_k}(\bm{\chi})$
directly by $P(\bm{\chi} \in C_k)$, which is provided by the classification algorithm.

Another solution overcoming the interface problem is to generate a specific model at the interface either with all the training data or with only a restriction.~\citet{Zhan2015} propose to enclosed the clusters by freezing the classifier and sampling the decision boundary. This strategy is very efficient for two-dimensional parameter spaces where the boundary decision is a line. However, in higher dimensions, the decision boundary is represented by a hyperplan. Its sampling at a low computational budget becomes very challenging. In addition, a clear definition of the boundary decision is not always available.

Choosing the best strategy is guided by several observations. First of all, the soft clustering mixes several physical regimes. It can lead potentially to nonphysical predictions and can amplify extrapolations of the reduced coordinates near boundary decisions, one of the main weakness of GPR. Then, building a model at the interface will adds complexity to the final surrogate model as extra parameters must be introduced to identify the points in the neighborhood. Moreover, even if a global model at the interface might be a possible solution it remains sensible to the definition of the region overlapping the boundary decision. Thus, the solution of a local interface model cannot be retained and a classical hard-split approach is preferred. 

\section{Improvement of the method}
\label{sec:improvement}
This section presents several improvements to the Local Decomposition Method introduced by~\citet{Dupuis2018b}. First, several methods are proposed to choose the number of clusters, based on Silhouette coefficient and local cross-validation. Then, a version of the adaptive sampling for parallel computing is presented and coupled with a local stopping criterion, independent for each cluster. Finally, as predicting values at the boundary between local models is very challenging, a specific data enrichment method near the boundary is introduced.

\subsection{Optimal number of clusters}
The unsupervised learning algorithm needs to set various numerical parameters in order to extract information from data. For instance, the number of clusters $K$  must be fixed for K-means. Various methods have been developed to determine these parameters but they are mainly heuristic and graphical. It is important to note these methods should be used as a guideline to estimate the numerical parameters and not as an absolute rule.

\subsubsection{Silhouette coefficient}
Two main properties define the quality of the clustering: the tightness and the separation. Ideally, a cluster must be tight, meaning the points belonging to a cluster are close together. The tightness can be measured for instance by the average distance between the points in the same cluster and the centroid of the cluster. The term separation means the points of a given cluster should be far from points in other clusters.

Silhouette analysis~\cite{Rousseeuw1987} refers to a method of interpretation and validation of consistency within clusters. It quantifies if clusters meet the requirements of tightness and separation. For a given sample $\bm{b_i}$ belonging to the cluster $C_k$, the value $\alpha(\bm{b_i})$ computes the average distance between $\bm{b_i}$ and all the other points in the same cluster $C_k$. This definition corresponds to the average dissimilarity of the cluster:
\begin{equation}
\alpha(\bm{b_i}) = \frac{1}{card(C_k) -1} \sum\limits_{\bm{b} \in C_k, \bm{b} \neq \bm{b_i} } dist( \bm{b_i}, \bm{b}).
\end{equation}
The second quantity $\gamma(\bm{b_i})$ selects the smallest value of $\alpha(\bm{b_i})$ if the cluster $C_k$ is substituted by a different one:
\begin{equation}
\gamma(\bm{b_i}) = \min\limits_{l \neq k} \frac{1}{card(C_l)} \sum\limits_{\bm{b}\in C_l} dist( \bm{b_i}, \bm{b}) .
\end{equation}
Finally, the combination of the two terms $\alpha$ and $\gamma$ defines the Silhouette coefficient $Sil$ as follows:
\begin{equation}
 Sil(\bm{b_i}) = 
\begin{cases}
     & 0, \text{ if }card(C_k) =1\\
  & \frac{\gamma(\bm{b_i}) - \alpha(\bm{b_i})}{ max(\alpha(\bm{b_i}), \gamma(\bm{b_i}))} , \text{ otherwise}
\end{cases}.
\label{eq:silhouette}
\end{equation}
The value of $Sil$ has a range of $[-1,1]$. A Silhouette coefficient close to $1$ indicates that the sample is far away from the neighboring clusters and that its cluster is homogeneous. A negative values can be interpreted as a wrong assignment of the sample to the cluster, while a value near $0$ can suggest a sample located between several clusters near the decision boundary. The Silhouette coefficient is computed for all the samples and then the average value is computed. The higher this value is, the more consistent the cluster is.

\subsubsection{Local cross-validation}
The cross-validation method can be used to determine the number of clusters.
This method, also called $K$-folds method, is a very popular approach to estimate the generalization error by using only the training set~\cite{Hastie2001}. Hence the surrogate model is trained with all the available data. The idea is to split the training set into $K$ disjoint subsets of approximately equal size. A subset is chosen as the testing set and the remaining subsets form the training set. The model is trained on this training set and evaluated on the testing set. This process is repeated $K$ times and the final error is computed as the average of the error one each subset.

A very specific case occurs when $K$ is chosen equal to $n$. This case is called leave-one-out cross validation. In our method, the leave-one-out predictivity coefficient is computed separately on each cluster to estimate the accuracy of each local model, so here $k$ is equal to $n_k$, with $n_k$ the number of samples belonging to the cluster $C_k$.
 Let consider $\bm{\widetilde{f}_{-i}}$ corresponds to the surrogate model based on the training sample $\bm{X_t}$ excluding $\bchi_i$. The leave-one-out error $e_{LOO}$ is written for the cluster $k$ as:
\begin{equation}
e_{LOO}^{(k)} = \frac{1}{n} \sum\limits_{i=1}^{n}\norm{\bm{f}(\bchi_i) - \bm{\widetilde{f}_{-i}(\bchi_i)}}^2, \ \forall \bm{\chi_i} \in C_k.
\end{equation}
From this expression, the leave-one-out predictivity coefficient $Q_2^{(k)}$ can be derived for the cluster $k$:
\begin{equation}
Q_2^{(k)} = 1 - \frac{e_{LOO}^{(k)}}{\widetilde{\sigma^{(k)}}^2},
\end{equation}
with $\sigma^{(k)}$ the standard deviation computed on the training sample belonging to the cluster $C_k$. If very low values are observed, the model can be considered as having poor predictive capabilities for this cluster. One can note that the leave-one-out method requires an additional computational endeavor. In the case of the ROMs, $n\times M$ models must be trained but this extra cost remains very low compare to a single CFD simulation.

\subsection{Improvement of the adaptive sampling}
\subsubsection{Batch distribution for parallel computing}
The active learning strategy presented in the previous section raises the following issue: should the simulations be computed sequentially or do parallel options exist? The sequential simulations allow to update the clustering and classification algorithms at each new simulation in order to have a very precise separation of the parameter space. However, with such an approach expensive high-fidelity models cannot be computed in parallel and all the available resources cannot be used. Therefore, a specific strategy has been developed to allow batch computation.

Let consider a batch of $n'$ simulations is performed. The objective is to distribute the computational budget on each cluster. The entropy can be measured locally on each cluster $k$, such as:
\begin{equation}
H_k = - \frac{1}{\log (|C_k|)}\sum_{i=1}^{n_k}p_i \log(p_i) \text{ with
} p_i = \frac{\lambda_i}{\sum\limits_{j \in C_k}\lambda_j},
\end{equation}
where $n_k$ is the number of training samples belonging to $C_k$.
A simple approach consisting of taking a ratio of entropy gives the number of new simulations $n_k'$ associated with the cluster $k$:
\begin{equation}
n_k' = \frac{H_k}{\sum\limits_{i=1}^K H_i} n'.
\end{equation}
\begin{figure}[ht]
\centering
\includegraphics[width=0.9\textwidth]{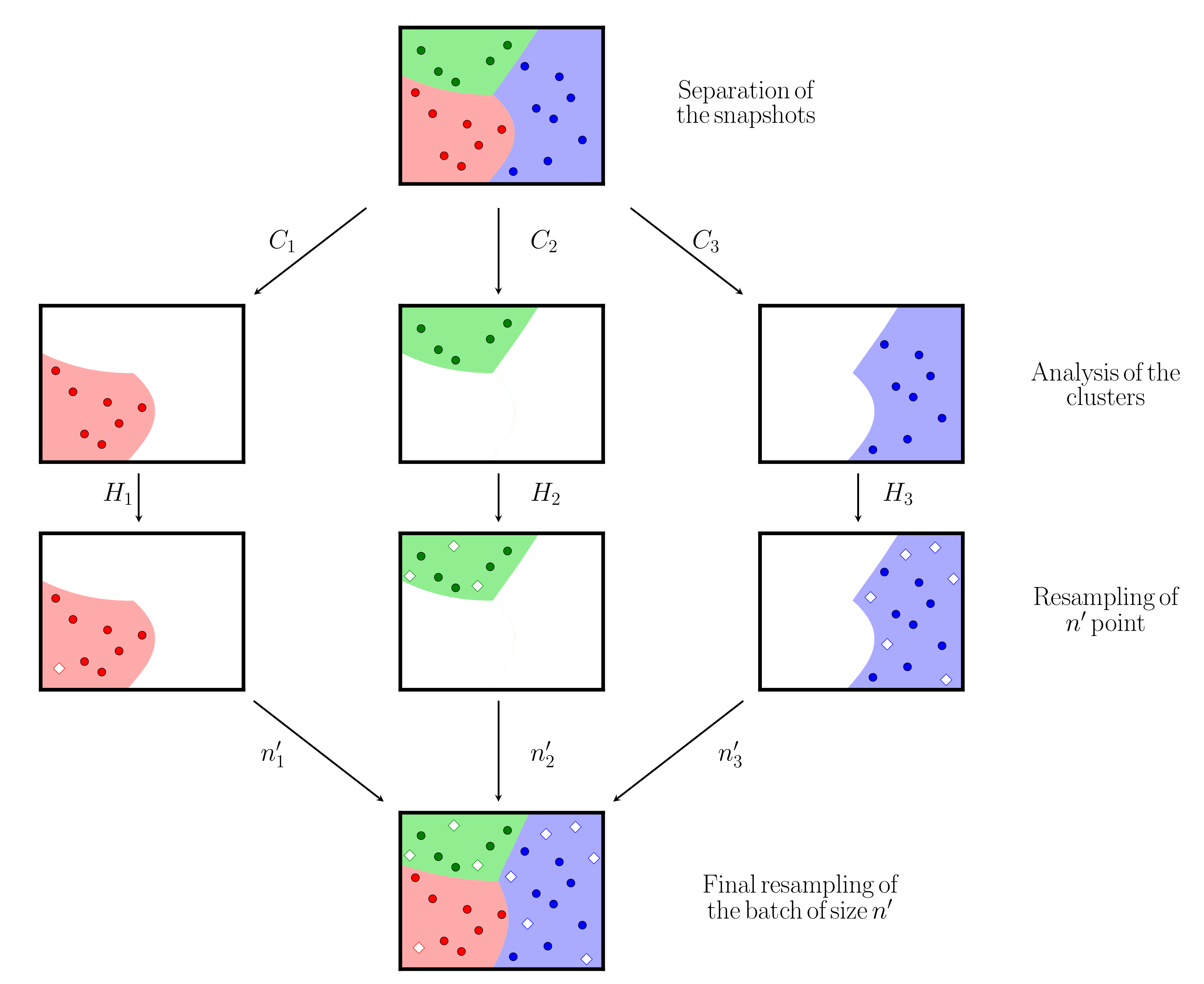}
\caption{Sketch of the LDM resampling process. New snapshots are marked with white diamonds.}
\label{fig:ldm_resampling}
\end{figure}
\autoref{fig:ldm_resampling} sketches the process of parallel resampling where $10$ new simulations are performed. In this example, after the analysis of the entropy, blue and green clusters benefit from the majority of the computational budget while the red cluster is considered as already well explored and does not require extra snapshots as its entropy is very low. Nevertheless, the size of the batch $n'$ must remain low compared to the global size of the training set in order to ensure that the cluster shapes and the classifier do not change too much. A classical ratio between $10\%$ and $20\%$ of the total size is usually a good trade-off.

\subsubsection{Local stop criterion}
This resampling strategy can be coupled with a stopping criterion adapted to the local properties of the LDM.
Instead of computing the predictivity coefficient $Q_2$~\cite{Iooss2010} by leave-one-out for all the snapshots, the $Q_2$ is computed and analyzed severally on each cluster. Thus, this criterion allows stopping the resampling for clusters of high quality if a minimum number of snapshots have been computed. The resampling can foster specific clusters with lower quality estimation and a significant amount of entropy.

This original approach ensures to spend the computational budget in the more challenging regions of the input parameter space. Nevertheless, determining the value of the threshold is not evident. It may depend on other clusters or on the initial samples. A classical value from experience is between $0.80$ and $0.99$. From our previous experiences with two-dimension and three-dimensional external aerodynamic cases, we fixed the threshold value to $0.95$.


\subsection{Data enrichment near the boundary}
In addition to the hard-split approach, a specific strategy improving predictions near decision boundaries has been developed. Indeed, the growing number of clusters increases directly the proportion of interface in the domain, raising problems of extrapolation in particular with the Gaussian Processes.

The previous method used a global ROM around the interface between two local models. However, the size of this intermediate region was difficult to determine and a global ROM could show high discrepancies in transition regions. The central idea here is to enrich each local model at the boundary with existing simulations. For each cluster, several snapshots located near the boundary do not belong to the cluster but can be useful to improve predictions near the interface. A least square problem is solved in order to determine values of reduced coordinates for points close to but outside the cluster. The strategy is divided into three steps repeated for all the clusters:
\begin{enumerate}
\item Finding the $r$ closest points to the cluster $k$ but not belonging to it. This set of points is written $\bm{X_r}$. The simplest method to find these points is to compute the distance between the center of the cluster and the other snapshots in the sensor space.
\item Determining reduced coordinates for the $r$ closest points solving the least square problem for the $k$-th cluster:
\begin{equation}
\min\limits_{a_j(\bm{\chi})} \norm{ \sum\limits_{j=1}^{M_k} a_j(\bm{\chi}) \bm{\phi_j} - \bm{f}(\bm{\chi}) }, \,  \bm{\chi} \in \bm{X_r},
\label{eq:ldm_proj_error}
\end{equation}
where $M_k$ denotes the number of POD modes for the $k$-th cluster.
\item Adding these new reduced coordinates to the training set of the cluster $k$. If the projection error of a given extra point does exceed a threshold, this point is rejected.
\end{enumerate}
The optimal value of $r$ can be determined by analyzing projection errors given in~\autoref{eq:ldm_proj_error}. Snapshots with important errors must not be added to the local training set as they can significantly decrease the accuracy of the local model. This strategy can be seen as regression methods overlapping other subspaces. POD modes are not updated with these new data in order to avoid contamination of other physical regimes, only the regression models are improved to reduce extrapolation errors.

\section{Application to industrial-like XRF-1 aircraft configuration}
\label{sec:XRF1}
The Airbus large transport aircraft of the eXternal Research Forum, also called the XRF-1, has been developed to
facilitate exchanges between academia and industry. It is used in this section in order to assess the LDM capabilities to
deal with industrial constraints and input data. The XRF-1 is a wing-body configuration, representative of modern civil
transport aircraft. It is designed to carry about 300 passengers with a maximum design range of $5,500$ nautical miles. This
reference configuration has already been used as a use case for the DLR project Digital-X~\cite{Kroll2016} and for the MDA-MDO
project~\cite{gazaix2019industrial}.

The XRF-$1$ presents a greater complexity than the AS$28$G configuration, previously studied in~\cite{Dupuis2018}, mainly due to a number of input variables leading to a large parameter space coupled with an important computational cost. Therefore, a deep study is dedicated to the sensibility of the parameter space decomposition against the number of training samples for $2$, $3$, and $4$ clusters. The primary goal is to understand the different clustering results and to interpret them. This study is focused on the wing in order to analyze deeply the regions encountered different physical regimes.

\subsection{Presentation of the case}
\subsubsection{Geometry and mesh}
\begin{figure}
\centering
\begin{subfigure}{1.0\textwidth}
\centering
   \includegraphics[width=0.5\linewidth]{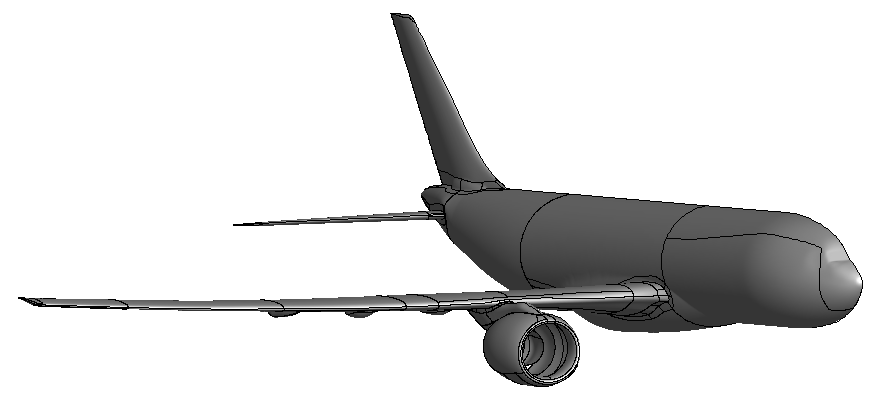}
\caption{Geometry.}
\label{fig:xrf1_geom}
\end{subfigure}
\begin{subfigure}{.49\textwidth}
\centering
   \includegraphics[width=1.0\linewidth]{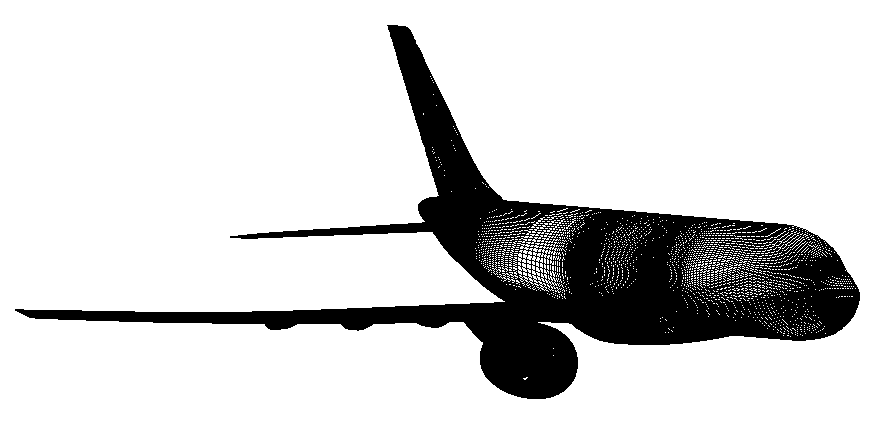}
\caption{Mesh.}
\label{fig:xrf1_mesh}
\end{subfigure}
\begin{subfigure}{.49\textwidth}
\centering
\includegraphics[width=0.8\linewidth, trim={0cm 1.2cm 0cm 1.8cm},clip]{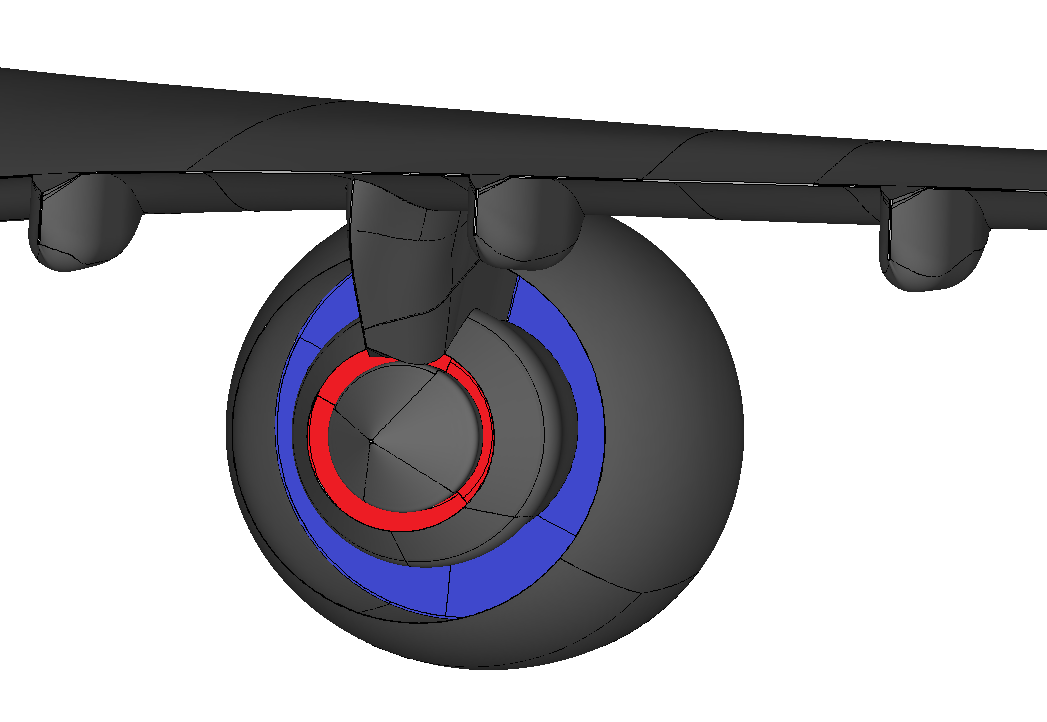}
\caption{Zoom on engine boundary conditions.}
\label{fig:xrf1_bc}
\end{subfigure}
\caption{XRF-1 aircraft configuration.}
\label{fig:xrf1_illus}
\end{figure}

A computer-aided design visualization of the XRF-1 configuration is shown in~\autoref{fig:xrf1_geom} and highlights the relative complexity of the geometry. Indeed, various systems are represented such as vertical and horizontal tail planes or the flap track fairings. This geometry is used for all flight phases. High lift devices are not considered with a view to simplification. Regarding the mesh, the XRF-1 configuration contains more than $120,000,000$ nodes for $324$ blocks, which is a lot more than the XRF-1 configuration in~\cite{dlr2016, Kroll2016}. The size of the domain is about $50$ chords. Specific wall treatment has been applied in order to have a dimensionless wall distance in the order of magnitude of the unity. Injection conditions are imposed on the engine exit face and on the fan exit face, depicted in~\autoref{fig:xrf1_bc} respectively in red and blue. An illustration of the surface mesh is given in~\autoref{fig:xrf1_mesh}. Moreover, only half aircraft is simulated as symmetric boundary conditions are used. The other boundaries of the computational domain are defined by the reference conditions of the upstream flow. To finish, adiabatic walls are considered.

\subsubsection{Solver settings}
\label{subsec:xrf1_solver}
The simulations are carried out using the cell-
centered finite-volume solver \textit{elsA}-ONERA~\cite{Cambier2013}, whose the co-owners are Airbus, Safran, and ONERA. It has been developed at ONERA and solves the compressible RANS equations on structured grids.
From the numerical point of view, space is discretized by a first order Roe's scheme to improve the stability of the computations. The time integration is performed with the backward Euler implicit scheme: the
algebraic system is linearized with the LU-SSOR implicit method. The turbulence modeling is ensured by the
model of Spalart and Allmaras and a multigrid strategy on three mesh levels is employed.

\subsubsection{Parameter space}
The study of the XRF-1 flight envelope involves seven different parameters: Mach
number, angle of attack, Reynolds number, two total pressures at the exit of the core, and two total temperatures at the exit of the fan. The stagnation quantities are used to defined injection boundary conditions using a characteristic relation and the Newton method.~\autoref{tab:xrf1_free_stream} sums up their variations which are representative of parametric studies in the aeronautical industry. The engine parameters are nondimensionalized.

A testing set of $190$ snapshots from a Sobol sequence~\cite{Bratley1988} has been built, corresponding to roughly $400,000$ CPU hours (on Intel 12 CPUs E5-2680-v3 at 2.5 Ghz). The parameter space is truncated in the region of high Mach numbers and high angles of attack. These points are not representative of a real flight mission and they encounter very important unsteadinesses.
\begin{table}[h!]
\centering
\begin{tabular}{cc}
     \hline\hline
Freestream variable& Amplitude of variation  \\\hline
Mach number & 0.1 - 0.87\\
Angle of attack ($^\circ$)&  0.0 - 6.0 \\
Reynolds number & $2.0 \times 10^6$ - $9.5 \times 10^7$ \\
$T_t$ core & 2.4 - 3.2 \\
$P_t$ core & 1.0 - 1.5\\
$T_t$ fan & 1.0 - 1.3\\
$P_t$ fan & 1.0 - 1.6\\
    \hline\hline
\end{tabular}
\caption{\label{tab:xrf1_free_stream} Freestream conditions.}
\end{table}

\subsection{Building surrogate models}
\subsubsection{Sampling}
\begin{figure}[ht]
\centering
   \begin{subfigure}{1.\textwidth}
   \begin{subfigure}{.5\textwidth}
  \centering
 \includegraphics[width=1.05\linewidth]{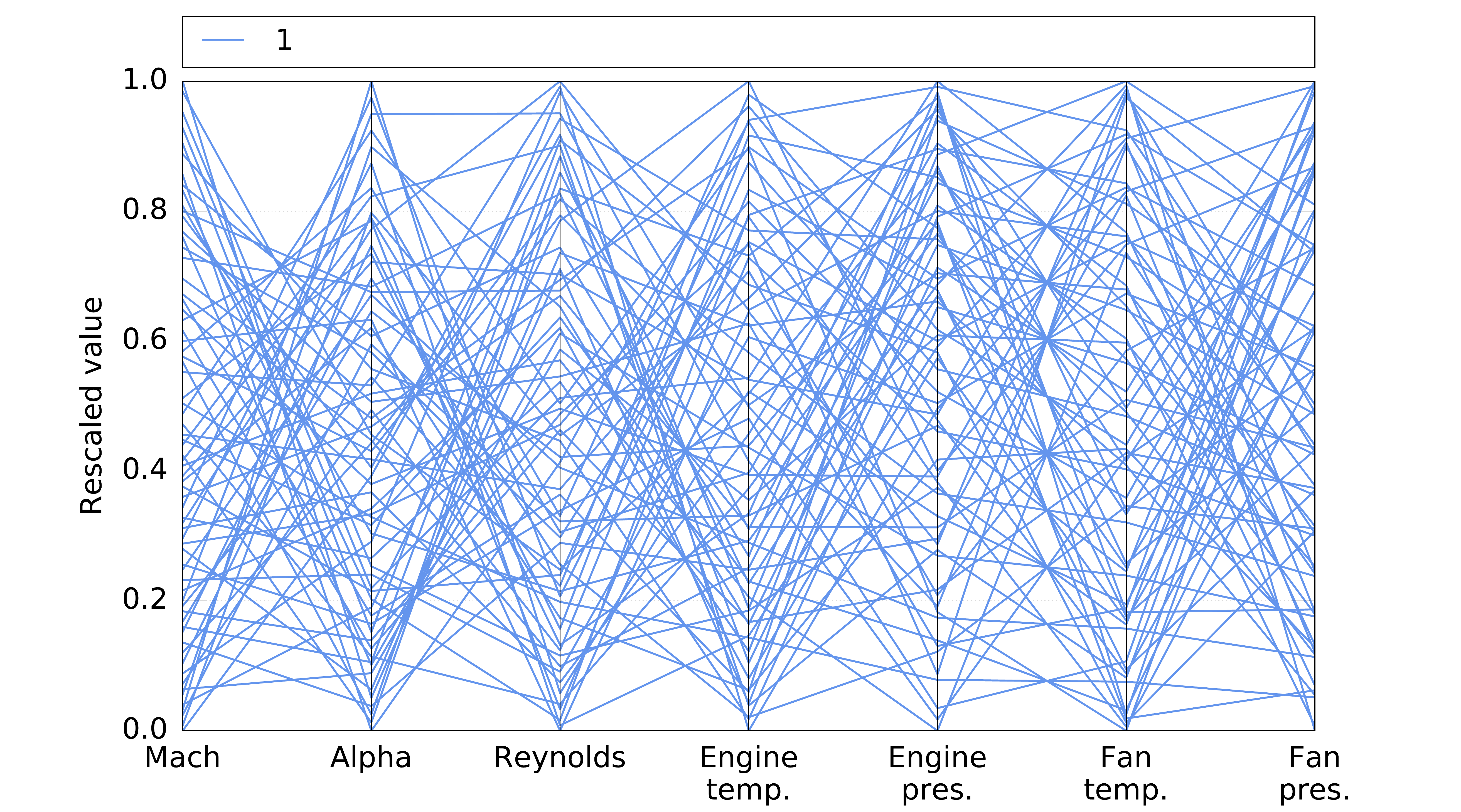}
  \caption{1 clusters.}
  \label{fig:xrf1_parallel_1}
  \end{subfigure}%
  \begin{subfigure}{.5\textwidth}
  \centering
 \includegraphics[width=1.05\linewidth]{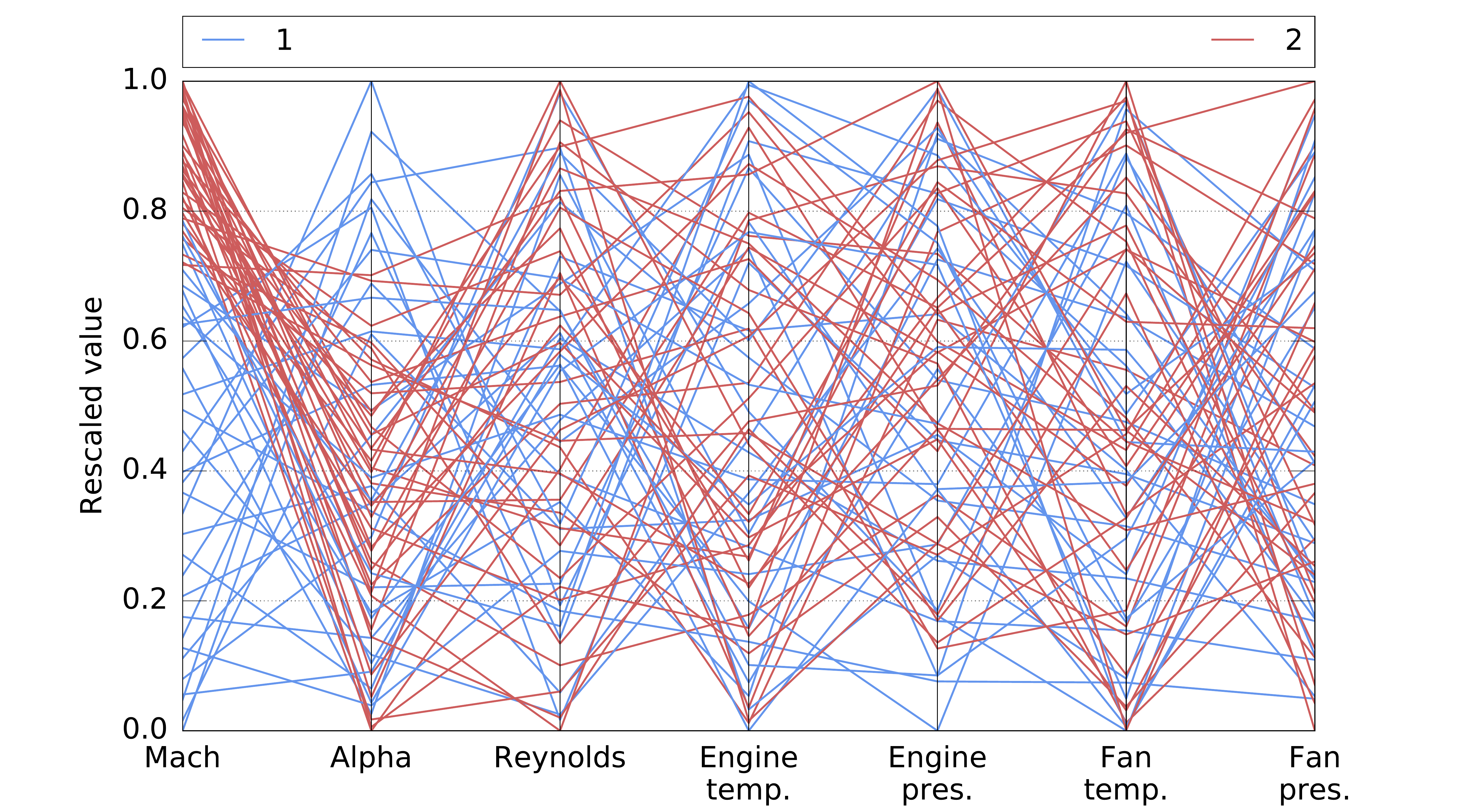}
  \caption{2 clusters.}
  \label{fig:xrf1_parallel_2}
  \end{subfigure}
\end{subfigure}
\begin{subfigure}{1.\textwidth}
   \begin{subfigure}{.5\textwidth}
  \centering
 \includegraphics[width=1.05\linewidth]{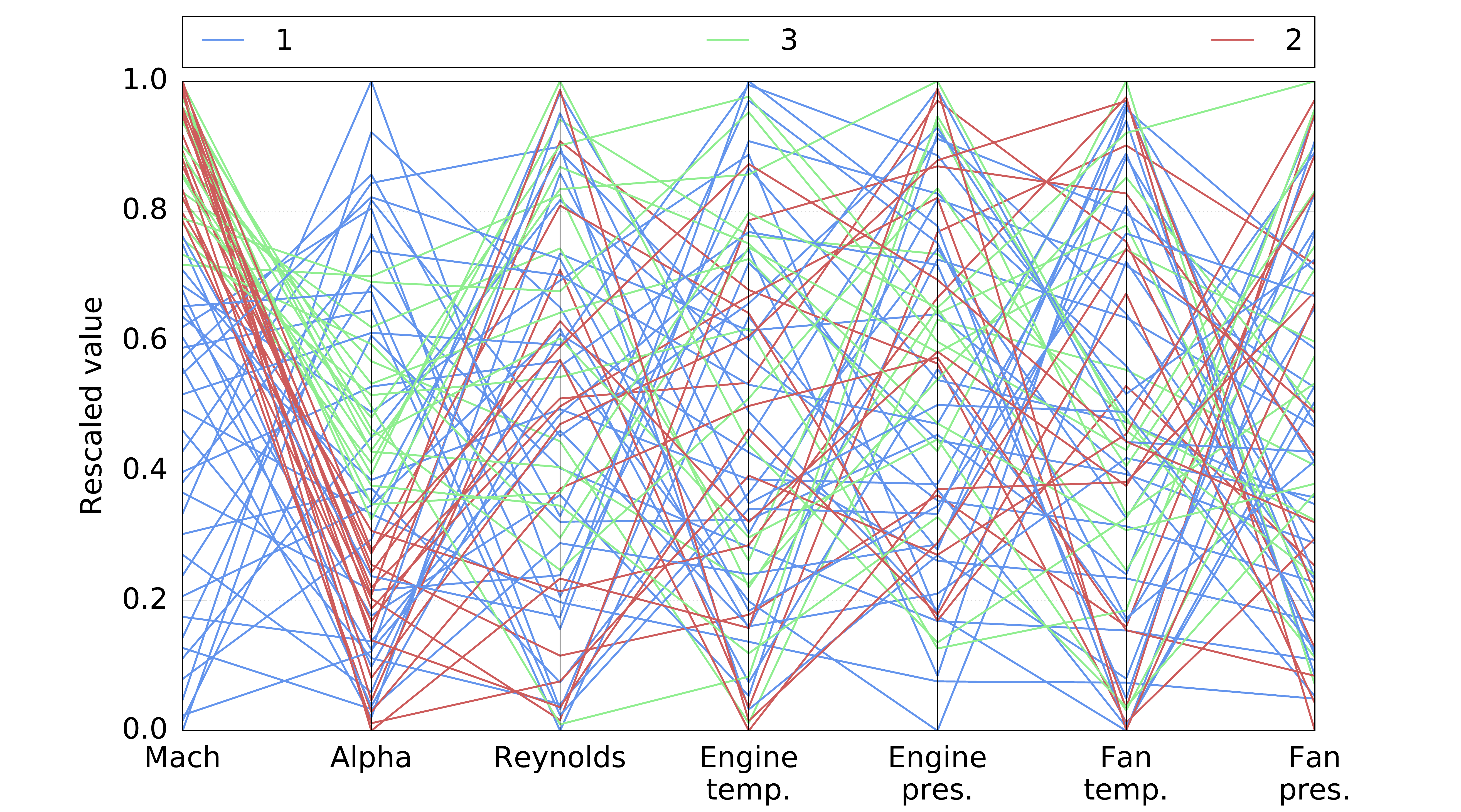}
  \caption{3 clusters.}
  \label{fig:xrf1_parallel_3}
  \end{subfigure}%
  \begin{subfigure}{.5\textwidth}
  \centering
 \includegraphics[width=1.05\linewidth]{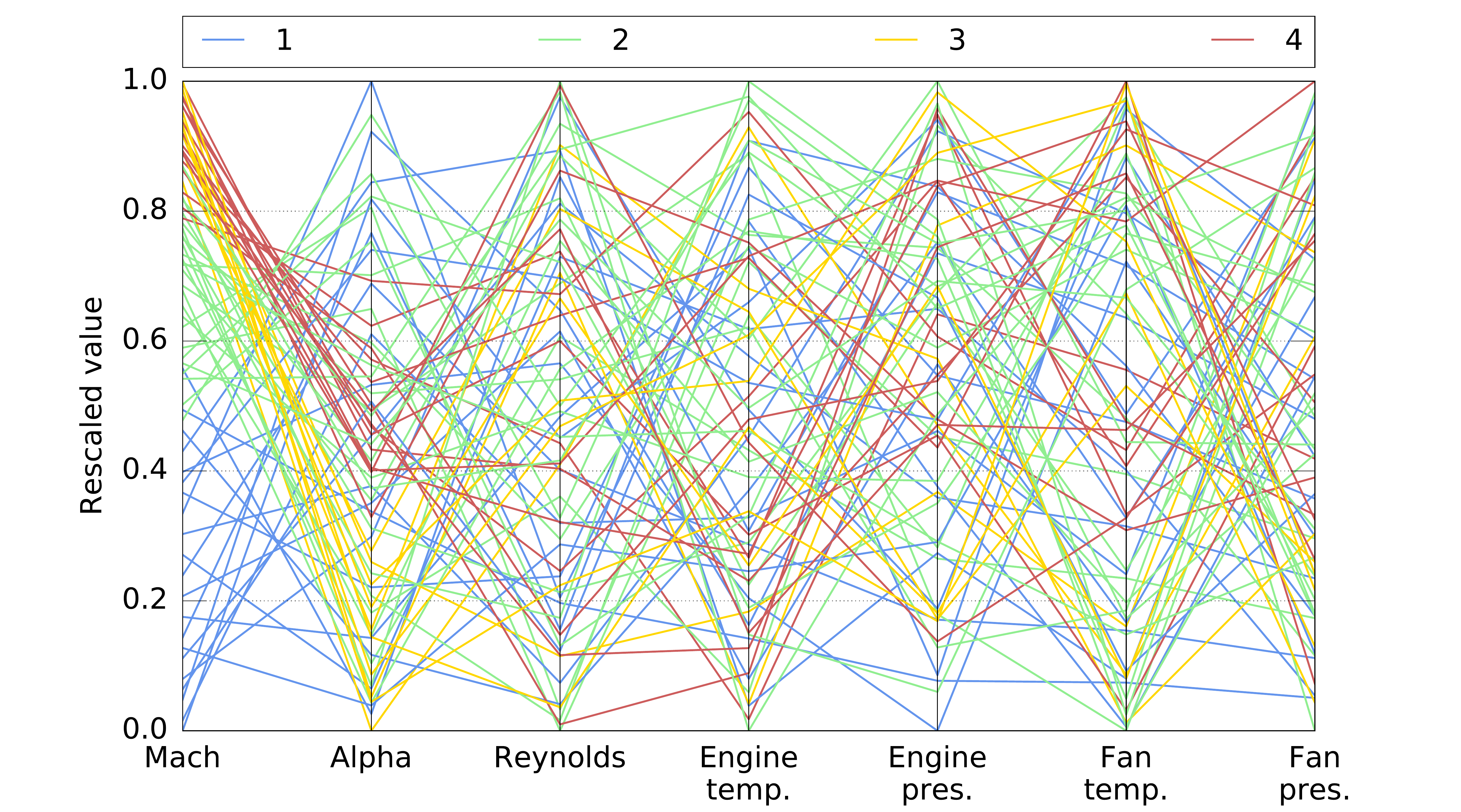}
  \caption{4 clusters.}
  \label{fig:xrf1_parallel_4}
  \end{subfigure}
\end{subfigure}
\caption{Distribution of the training samples in a parallel coordinate formalism. The belonging to the different cluster is represented in color.}
\label{fig:xrf1_parallel} 
\end{figure}

Final surrogate models are built from $70$ training samples, following the rule of thumb $10d$~\cite{Jones1998, Loeppky2009}.~\autoref{fig:xrf1_parallel} depicts the distribution of the training samples in a parallel coordinate formalism for the global approach and the LDM with $2$, $3$, and $4$ clusters. It includes the resampling for the LDM with $20$ initial simulations and the rest for the resampling. The case with $1$ cluster corresponds to a classical ROM without resampling. The most striking feature is the correlation between the Mach number and the clustering. By looking at~\autoref{fig:xrf1_parallel_2},~\ref{fig:xrf1_parallel_3}, and~\ref{fig:xrf1_parallel_4}, a clear clustering is observed on the Mach number axis, which is not the case for the other axes with the exception of the angle of attack. Indeed, in clusters can also be noticed on the angle of attack axis. As a consequence, the following analysis pays special attention to the Mach number and the angle of attack as other inputs seems to have a weak influence of the separation into several clusters.

 Comparing~\autoref{fig:xrf1_parallel_1},~\ref{fig:xrf1_parallel_2},~\ref{fig:xrf1_parallel_3}, and~\ref{fig:xrf1_parallel_4} highlights that the resampling focuses on regions of high Mach numbers, which is a result similar to RAE$2822$ and AS$28$G~\cite{Dupuis2018b, Dupuis2018}. Indeed, the sampling of the LDM, irrespective of the number of clusters, is much denser above a rescaled Mach number of $0.7$ (a real Mach number of about $0.6$), while on the contrary the sampling below a value of $0.7$ is quite sparse compared to the classical method. Therefore, the LDM method focuses on the transonic regime transition and on transition regions between subsonic and transonic.
\begin{figure}
\centering
   \begin{subfigure}{1.\textwidth}
   \begin{subfigure}{.5\textwidth}
	\centering  
	\includegraphics[width=0.9\linewidth, trim={0cm 0.2cm 0.2cm 0.2cm},clip]{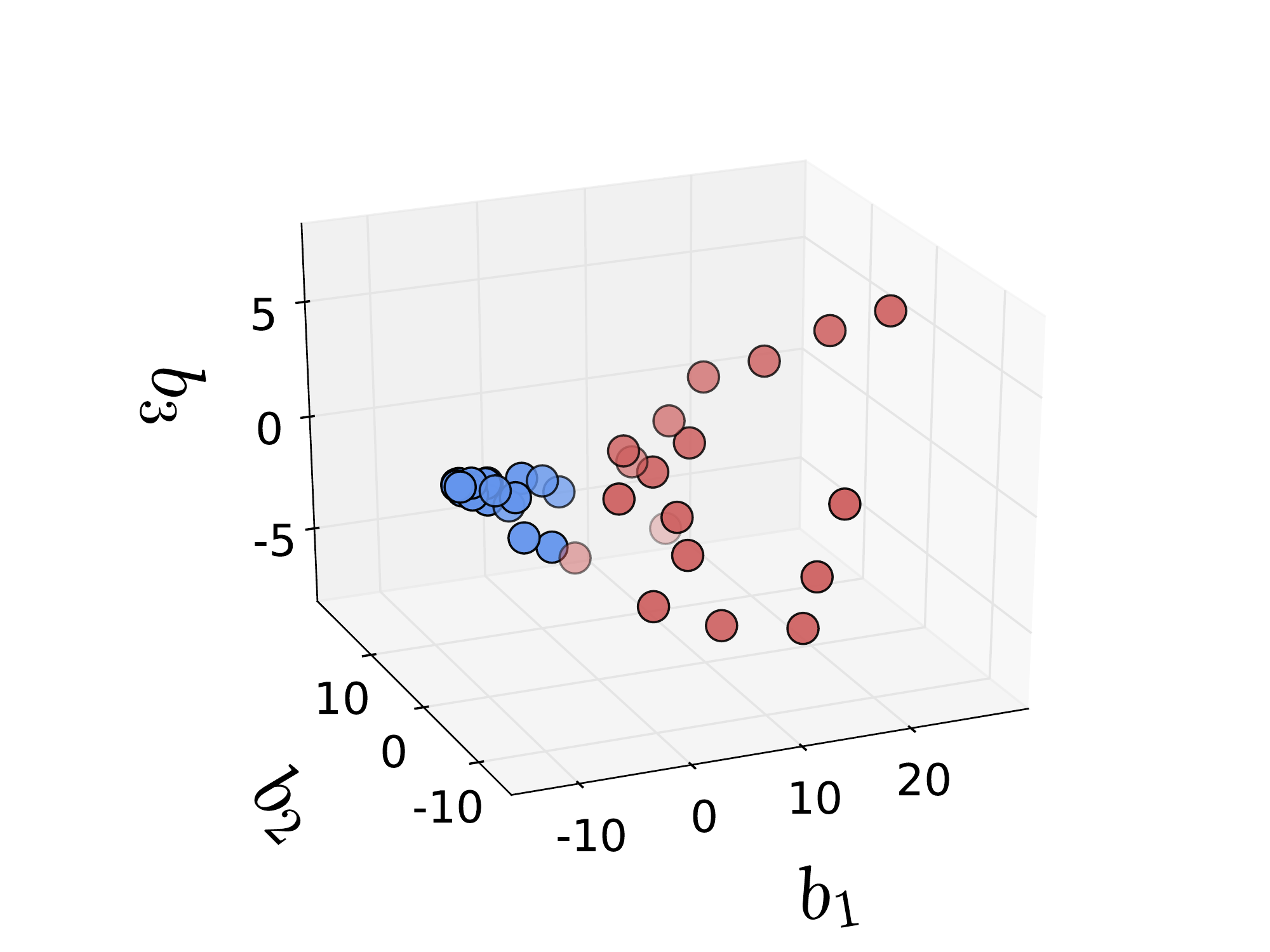} 
	\caption{40 training samples.}  
  \end{subfigure}%
  \begin{subfigure}{.5\textwidth}
	\centering  
	\includegraphics[width=0.9\linewidth, trim={0cm 0.2cm 0.2cm 0.2cm},clip]{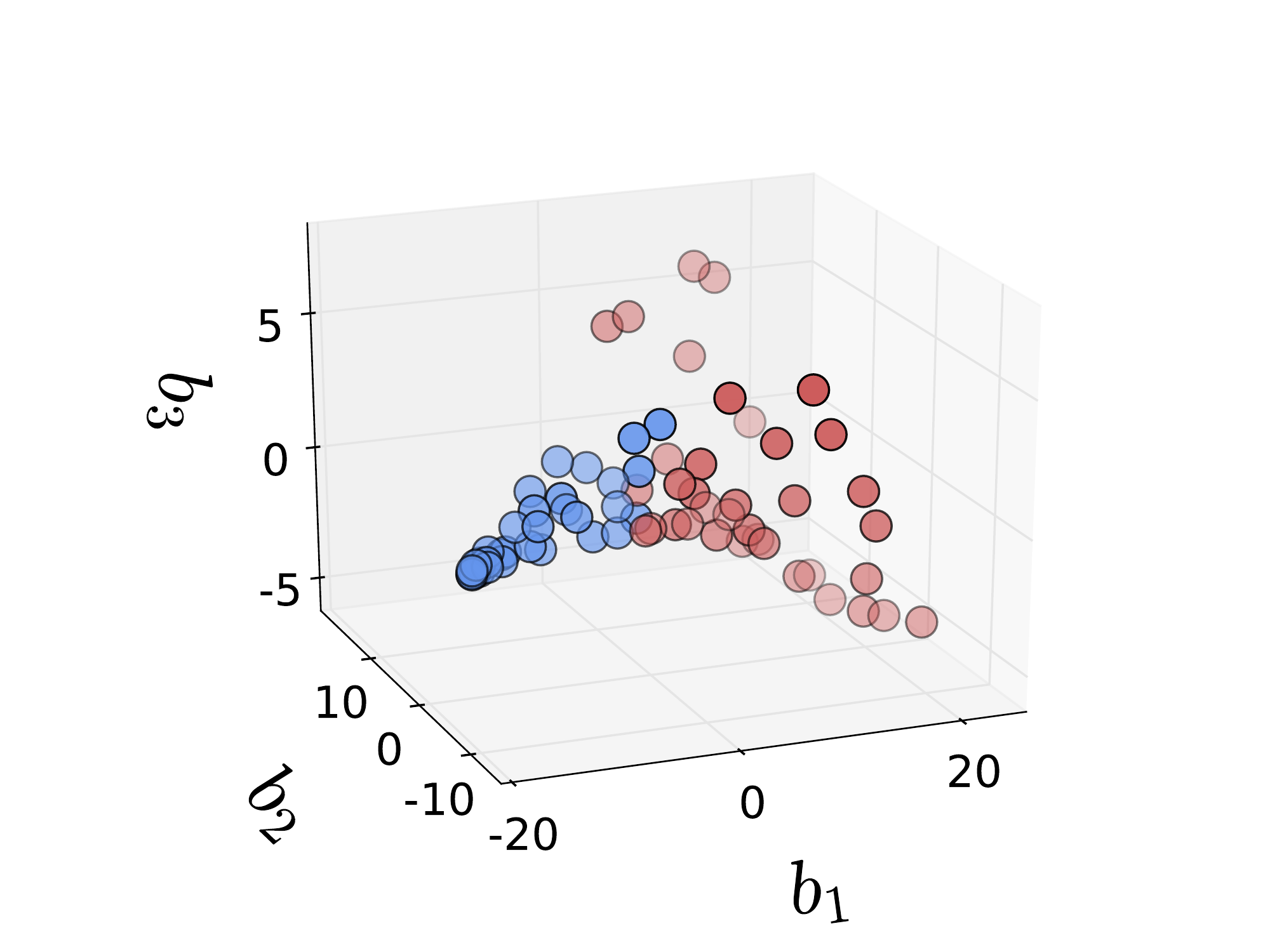} 
	\caption{70 training samples}
  \end{subfigure}
   \caption*{2 clusters}
\end{subfigure}
\begin{subfigure}{1.\textwidth}
   \begin{subfigure}{.5\textwidth}
	\centering  
	\includegraphics[width=0.9\linewidth, trim={0cm 0.2cm 0.2cm 0.2cm},clip]{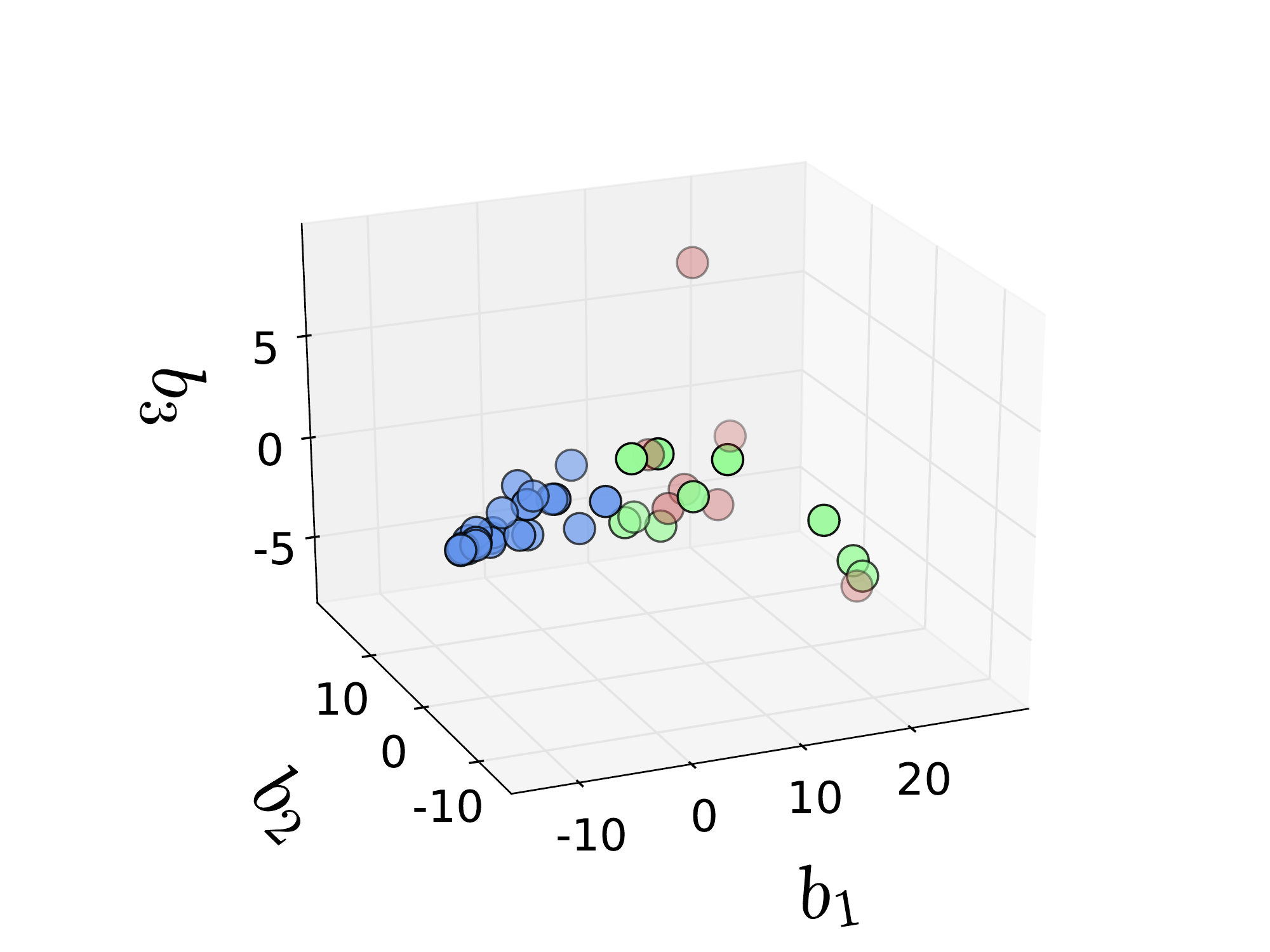}
	\caption{40 training samples for $3$ clusters}  
  \end{subfigure}%
  \begin{subfigure}{.5\textwidth}
	\centering  
	\includegraphics[width=0.9\linewidth, trim={0cm 0.2cm 0.2cm 0.2cm},clip]{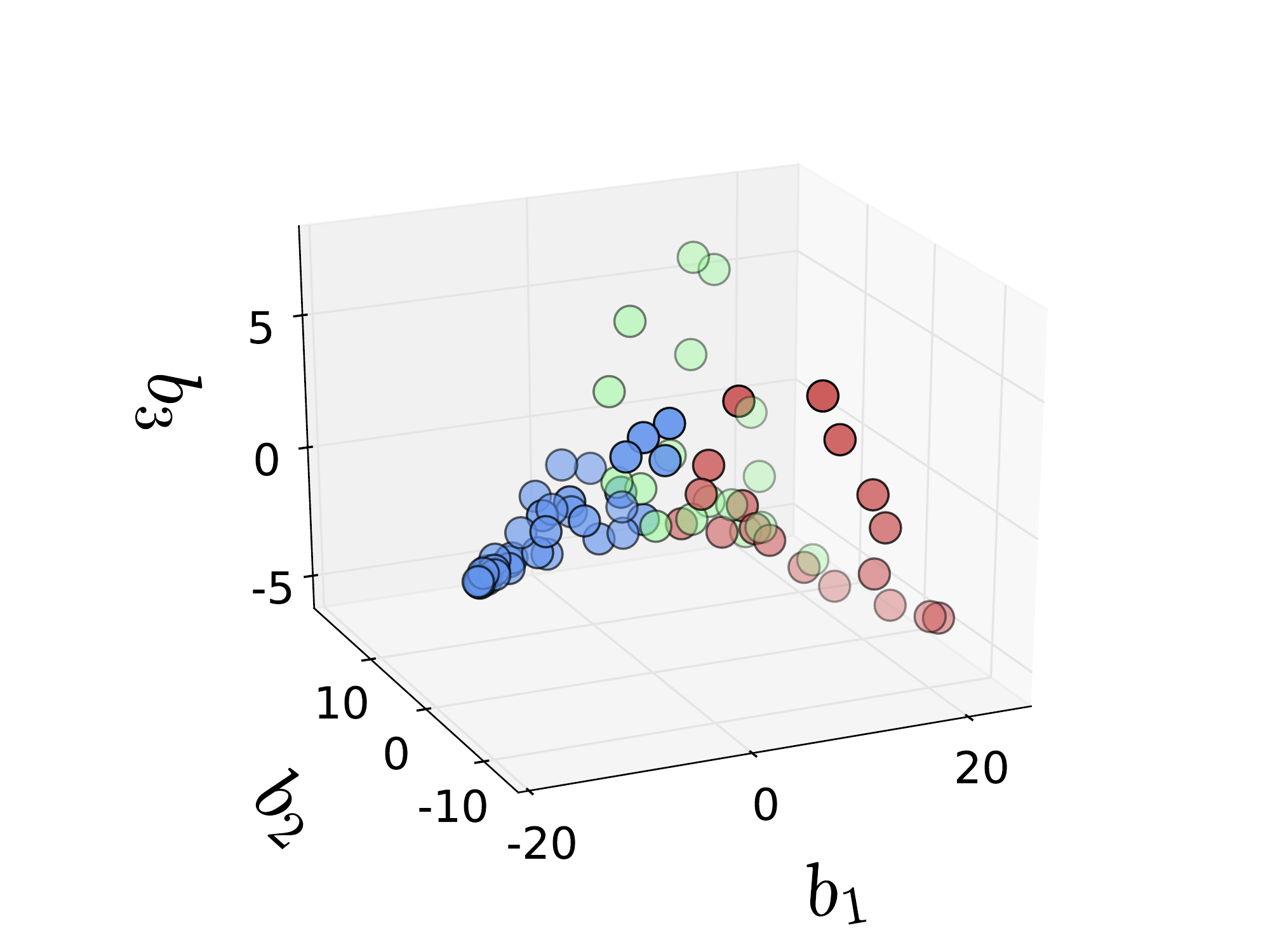}
	\caption{70 training samples for $3$ clusters}
  \end{subfigure}
   \caption*{3 clusters}
\end{subfigure}
\begin{subfigure}{1.\textwidth}
   \begin{subfigure}{.5\textwidth}
	\centering  
	\includegraphics[width=0.9\linewidth, trim={0cm 0.2cm 0.2cm 0.2cm},clip]{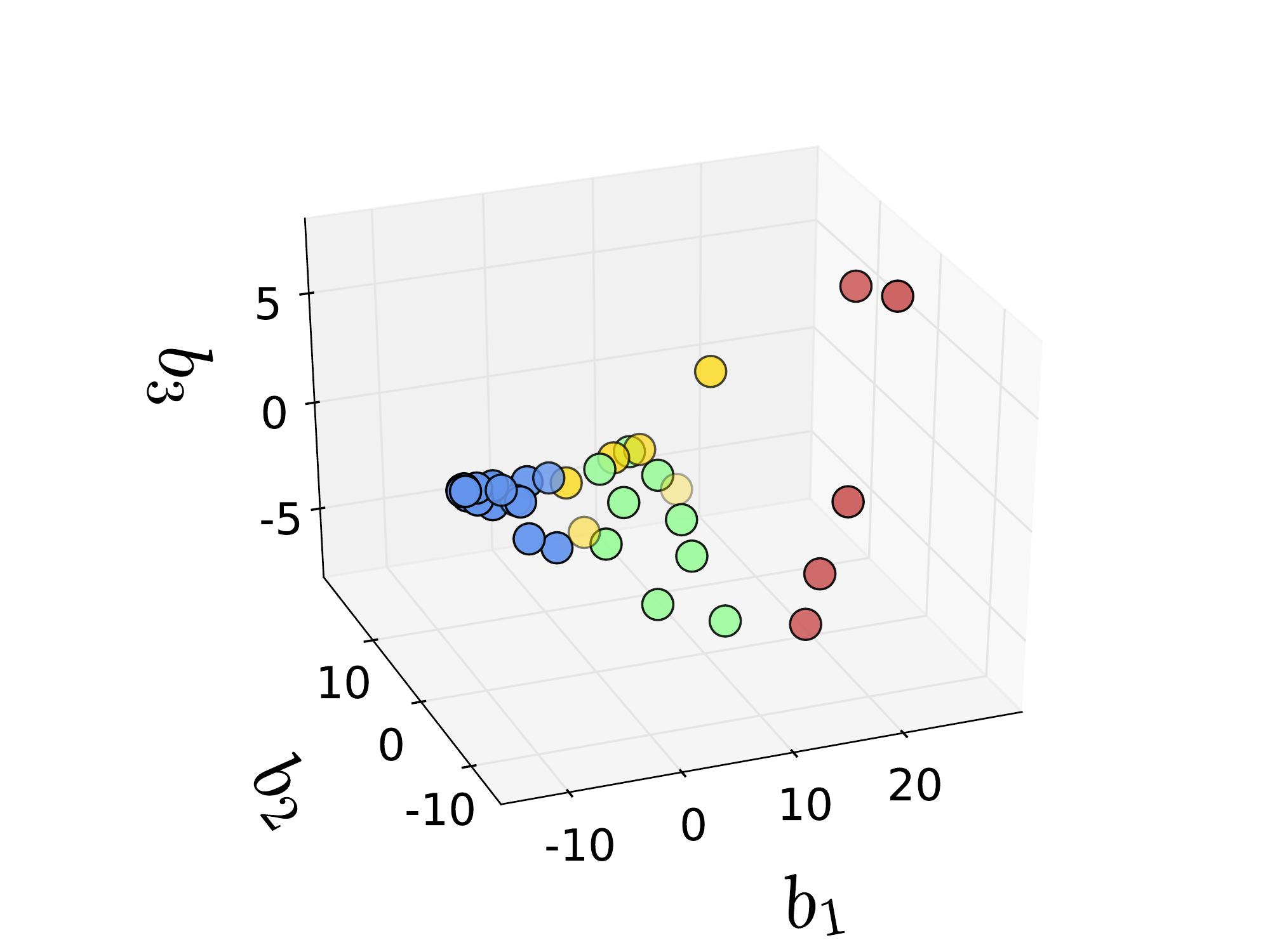}
	\caption{40 training samples for $4$ clusters}  
  \end{subfigure}%
  \begin{subfigure}{.5\textwidth}
	\centering  
	\includegraphics[width=0.9\linewidth, trim={0cm 0.2cm 0.2cm 0.2cm},clip]{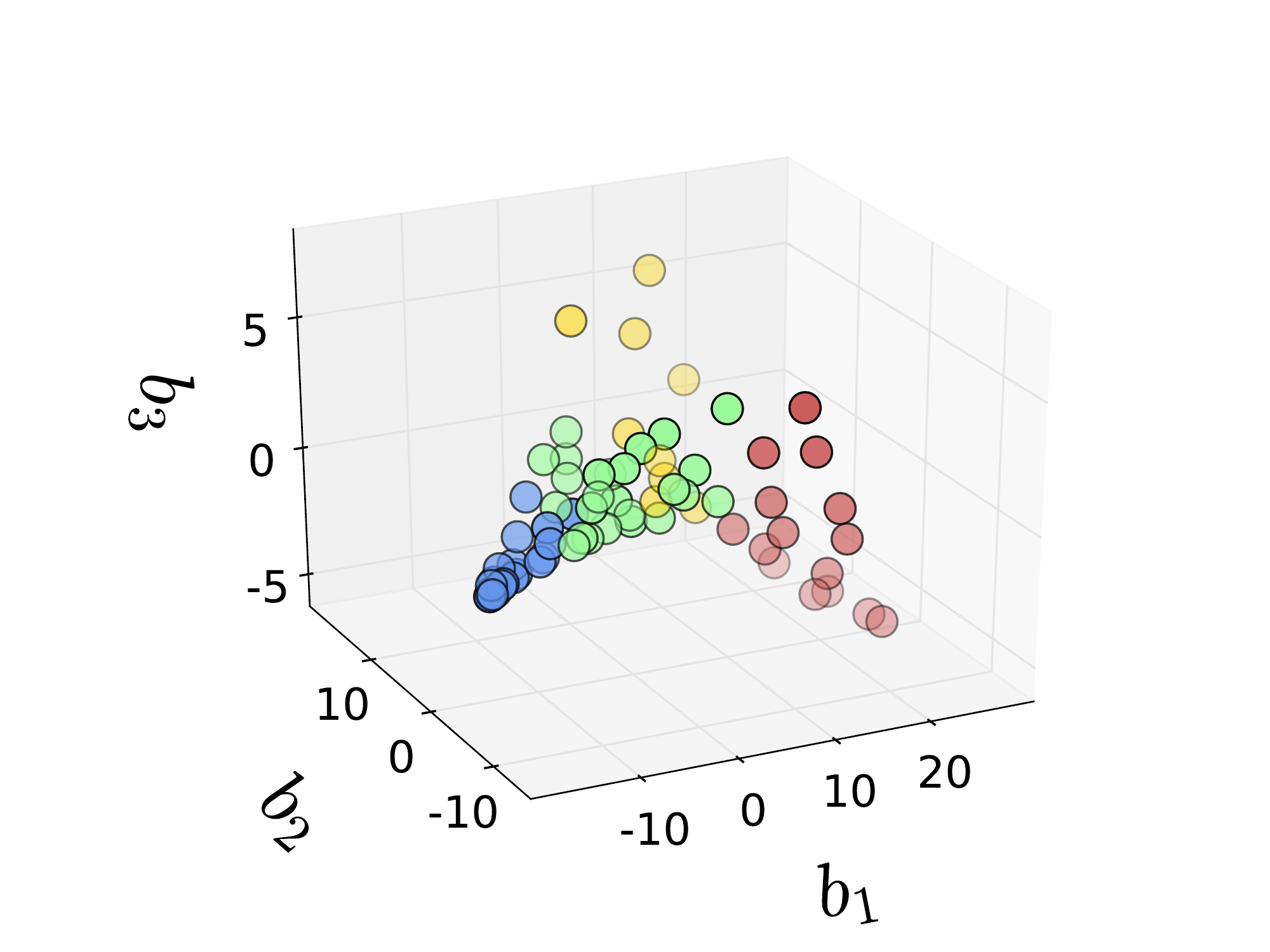}
	\caption{70 training samples for $4$ clusters}
  \end{subfigure}
   \caption*{4 clusters}
\end{subfigure}
\caption{Clustering with different numbers of clusters and training samples. $b_1$, $b_2$, and $b_3$ correspond to the first three coordinate of the reduced space for the shock sensor. }
\label{fig:xrf1_clustering}
\end{figure}

\subsubsection{Clustering and classification}
This section aims at analyzing and comparing the behavior of the LDM for different numbers of clusters. The $K$-means algorithm appears to be a suitable choice as it ensures quite homogeneous clusters, even for an important number of clusters. The DBSCAN method is limited to $2$ clusters by fixing the numerical parameter with the Elbow method while the GMM was less stable.

\autoref{fig:xrf1_clustering} depicts the clustering of the shock sensor for different numbers of clusters and two numbers of training samples. A clear pattern is observed with $40$ training samples: some clusters are concentrated in a specific region (the blue dots) while the others are spread across a manifold. Thus, the clustering of the spread snapshots gives quite sparse clusters. Nevertheless, the resampling corrects the problem by adding information in these clusters, as shown by the clustering with $70$ training samples. A clear symmetric shape according to the $b_2$ axis (around $b_2 = 0$) can be observed. Each specific number of clusters gives a different interpretation to this distribution of samples within the reduced space of the shock sensor. Computing average Silhouette coefficients shows very values for all configurations: $0.437$ for $2$ clusters, $0.419$ for three clusters, and $0.411$ for three clusters. Therefore, these very small differences do not allow to conclude on the best number of clusters and the best interpretation given to the data. A broader analysis must be undertaken.

\autoref{fig:xrf1_cluster_vs_N} illustrates the classification of the parameter space for different numbers of clusters and training samples. Its analysis aims to improve the understanding of the method. The case with $4$ clusters and $30$ training samples is not represented as some clusters have insufficient size. First, it can be noticed that the classification remains quite stable for $2$ and $4$ clusters. On the contrary, the decomposition of the parameter space for $3$ clusters changes significantly as the number of training samples increases. In particular, the green and red clusters associated to high Mach numbers constantly change. Clear size and location are not identified.
As regards the particular classifications with $70$ training samples, several observations are proposed:
\begin{itemize}%
    \setlength{\itemsep}{5pt}%
    \setlength{\parskip}{0pt}%
\item a region of low Mach number, represented in blue, is clearly founded for all the models. Only the location of the boundary changes;
\item the classification given by the configuration with $2$ clusters is consistent with RAE$2822$ and AS$28$G results~\cite{Dupuis2018, Dupuis2018b}, as a small cluster of high Mach numbers is associated to a large cluster of smaller Mach numbers, while a linear dependency to the angle of attack is observed;
\item configurations with $3$ and $4$ clusters show similarities. The high Mach numbers are split along the values of the angle of attack;
\item the configuration with $4$ clusters reduces the size of the cluster $\#1$ in favor of an intermediate region.
\end{itemize}
These results highlight the challenge of correctly setting the method, as various interpretations can be proposed for this complex application. 

A specific interpretation of the reduced space for the shock sensor can be given by comparing cases with $70$ samples in \autoref{fig:xrf1_clustering} and \autoref{fig:xrf1_cluster_vs_N}. The first coordinate $b_1$ seems to be directly related to the Mach Number as evidenced by the clear separation between the blue cluster (for low Mach number) and the other ones along $b_1$. The two other coordinates $b_2$ and $b_3$ seems to be related to the angle of attack as they allow to split the data into three or four clusters.

\pagebreak[4]
\global\pdfpageattr\expandafter{\the\pdfpageattr/Rotate 90}
\begin{landscape}
    \vspace*{\fill}
\begin{figure}[h!]
\centering
\includegraphics[width=1.45\textwidth]{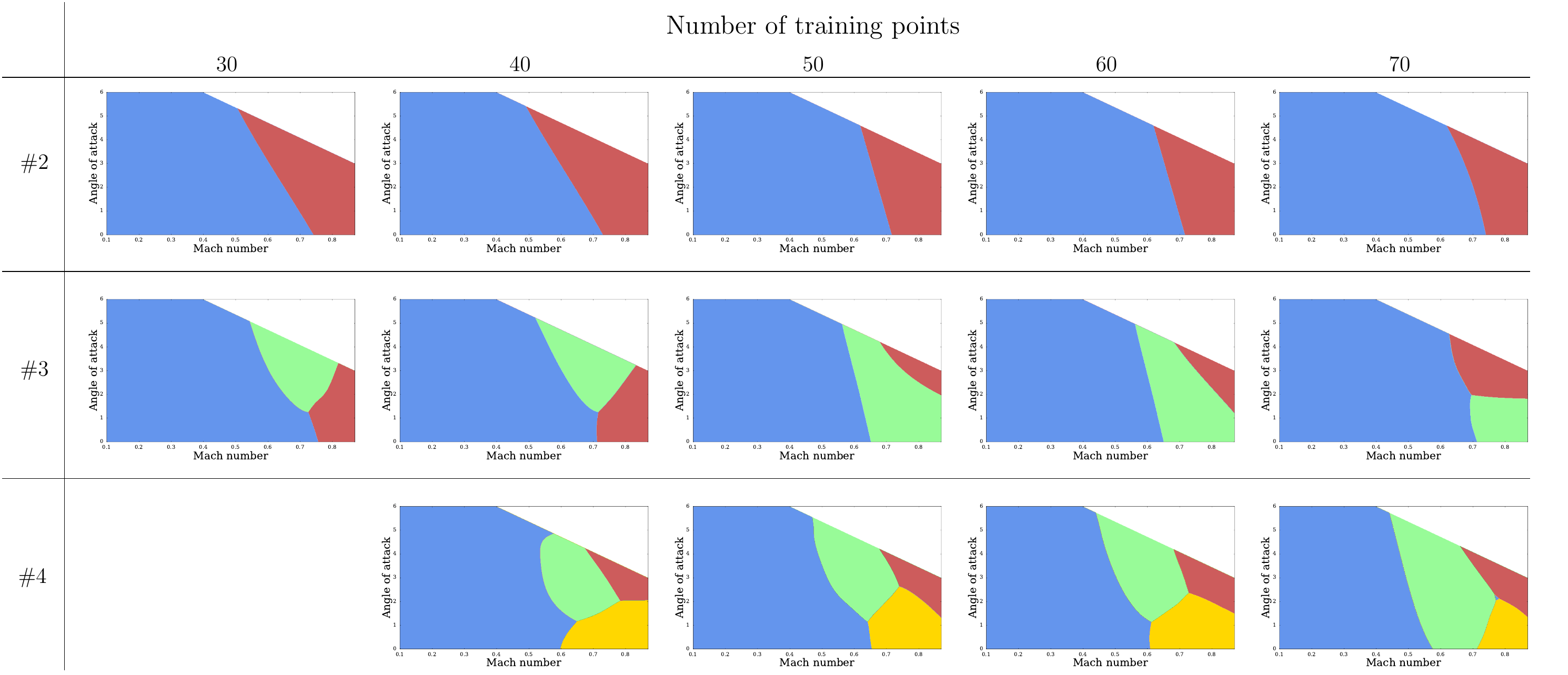}
\caption{Evolution of the parameter space decomposition for different number of training samples and different number of clusters.}
\label{fig:xrf1_cluster_vs_N}
\end{figure}
    \vspace*{\fill}

\end{landscape}
\pagebreak[4]
\global\pdfpageattr\expandafter{\the\pdfpageattr/Rotate 0}

\begin{figure}[h!]
\centering
   \begin{subfigure}{1.\textwidth}
   \begin{subfigure}{.5\textwidth}
	\centering  
	\includegraphics[width=1.1\linewidth]{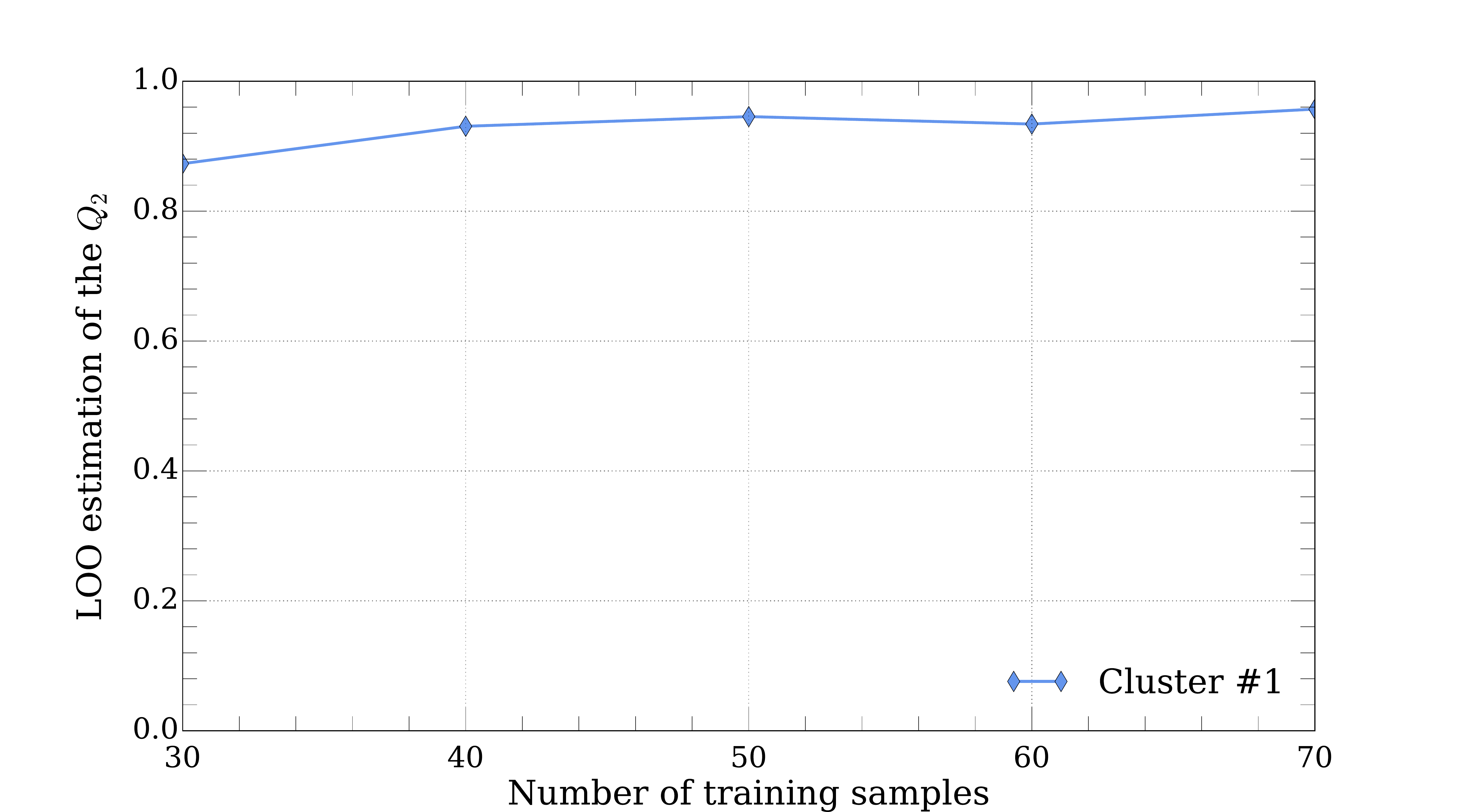}
	\caption{1 cluster}  
  \end{subfigure}%
  \begin{subfigure}{.5\textwidth}
	\centering  
	\includegraphics[width=1.1\linewidth]{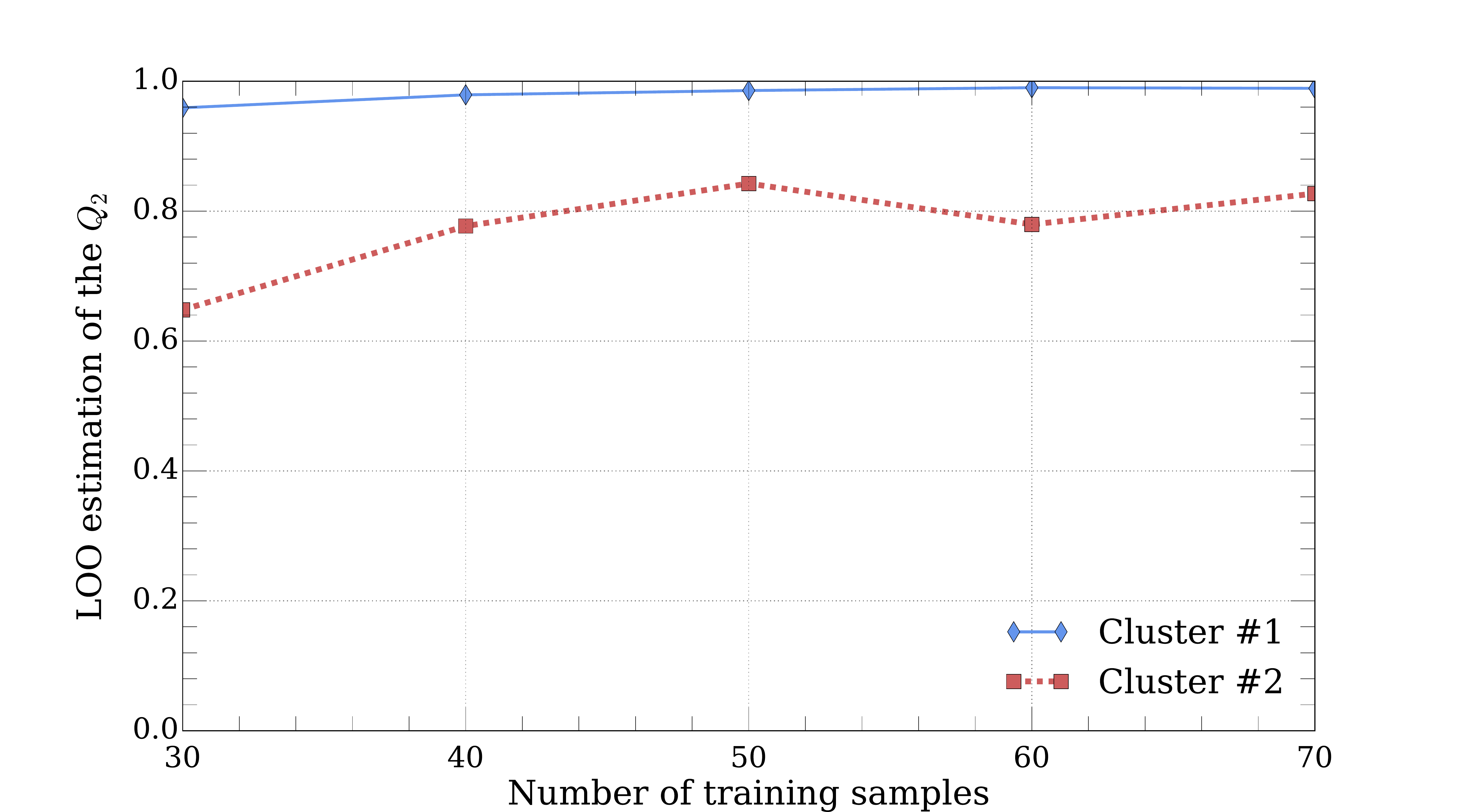} 
	\caption{2 clusters}
  \end{subfigure}
\end{subfigure}

\begin{subfigure}{1.\textwidth}
   \begin{subfigure}{.5\textwidth}
	\centering  
	\includegraphics[width=1.1\linewidth]{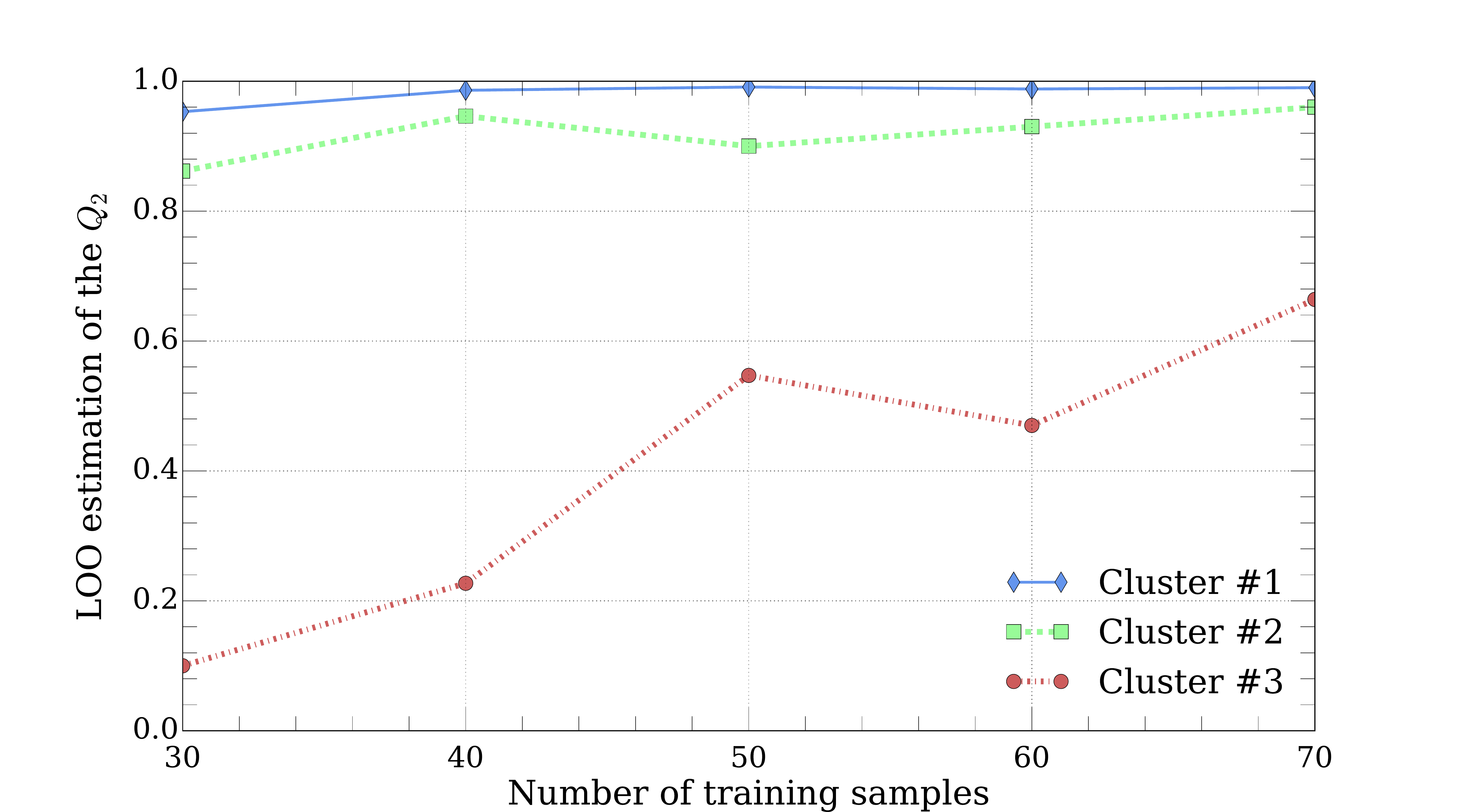}
	\caption{3 clusters}  
  \end{subfigure}%
  \begin{subfigure}{.5\textwidth}
	\centering  
	\includegraphics[width=1.1\linewidth]{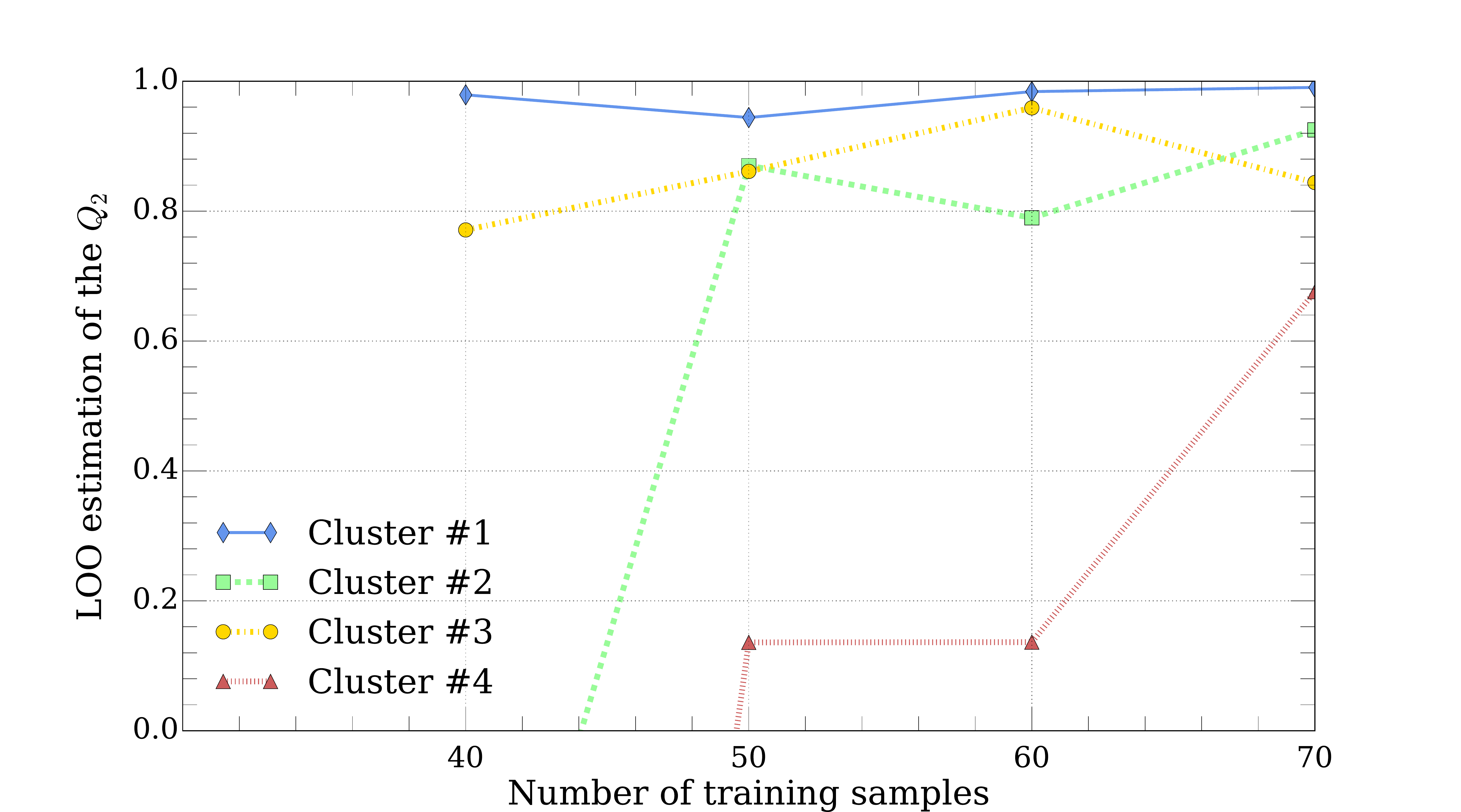} 
	\caption{4 cluster}
  \end{subfigure}
\end{subfigure}
\caption[Two numerical solutions]{Evolution of the leave-one out error.}
\label{fig:xrf1_loo}
\end{figure}

Cross-validation errors can help to obtain a reliable model and can guide in the choice of the best number of clusters.~\autoref{fig:xrf1_loo} provides the LOO error for different models. The models with $3$ and $4$ clusters show weaknesses when they are trained with a low number of training samples. In particular, some local models have negative or almost null values of $Q_2$ with $30$ and $40$ training samples. This result could be expected, as the number of training samples is not enough to sufficiently train each local model. It would not be appropriate to consider these models with such LOO results. One can notice that the predictivity coefficient reaches an asymptotic behavior for $1$ cluster from $30$ to $70$ training samples. As it will be seen later, the classical ROM does not show great performance between $30$ and $50$ samples. Therefore, this result calls into question the use of LOO estimation error for a single model with very different physical regimes. Indeed, a local analysis according to the specificity of the flow seems more relevant for the a-priori assessment of the ROM. This statement will be reinforced by the accuracy performance of the local ROMs. 

\subsubsection{POD analysis}
The analysis of the POD is only performed for $70$ training samples. The results are summarized in~\autoref{table:clustering_dr}. Overall, they are similar to those of the other aerodynamic cases~\cite{Dupuis2018, Dupuis2018b}. The cluster associated to low Mach numbers shows a low entropy compared to the other clusters. However, specific elements can be underlined: the clusters $\#2$ and $\#3$ of the model with $3$ clusters and the clusters $\#3$ and $\#4$ of the model with $4$ clusters are of the same order of magnitude with a high entropy. It might be thought that the Mach number is the only parameter responsible for the generation of entropy, as these clusters differ mainly by the angle of attack.
\begin{table}[ht]
\centering
\begin{tabular}{ccccc}
\hline\hline
Number of cluster & Cluster index& Snapshots& Modes& Entropy
\\\hline
1 & \#1 & 70& 32 & 0.481\\
\hline
\multirow{2}{*}{2} & \#1 & 38 & 22 & 0.441\\
& \#2 & 32 & 28 & 0.699\\
\hline
\multirow {3}{*}{3} & \#1 & 35& 20& 0.444\\
&\#2& 17& 15& 0.7397 \\
&\#3& 18&16&0.7287\\
\hline
\multirow{4}{*}{4}  & \#1& 21& 12& 0.466\\ 
 & \#2 & 26& 20& 0.542\\ 
 & \#3 & 10& 9& 0.750\\ 
 & \#4 & 15& 14& 0.766\\ 
 \hline\hline
\end{tabular}
\caption{\label{tab:xrf1_pod} Dimension reduction behavior.}
\label{table:clustering_dr}
\end{table}

\begin{landscape}
\begin{figure}[h!]
\includegraphics[width=1.45\textwidth]{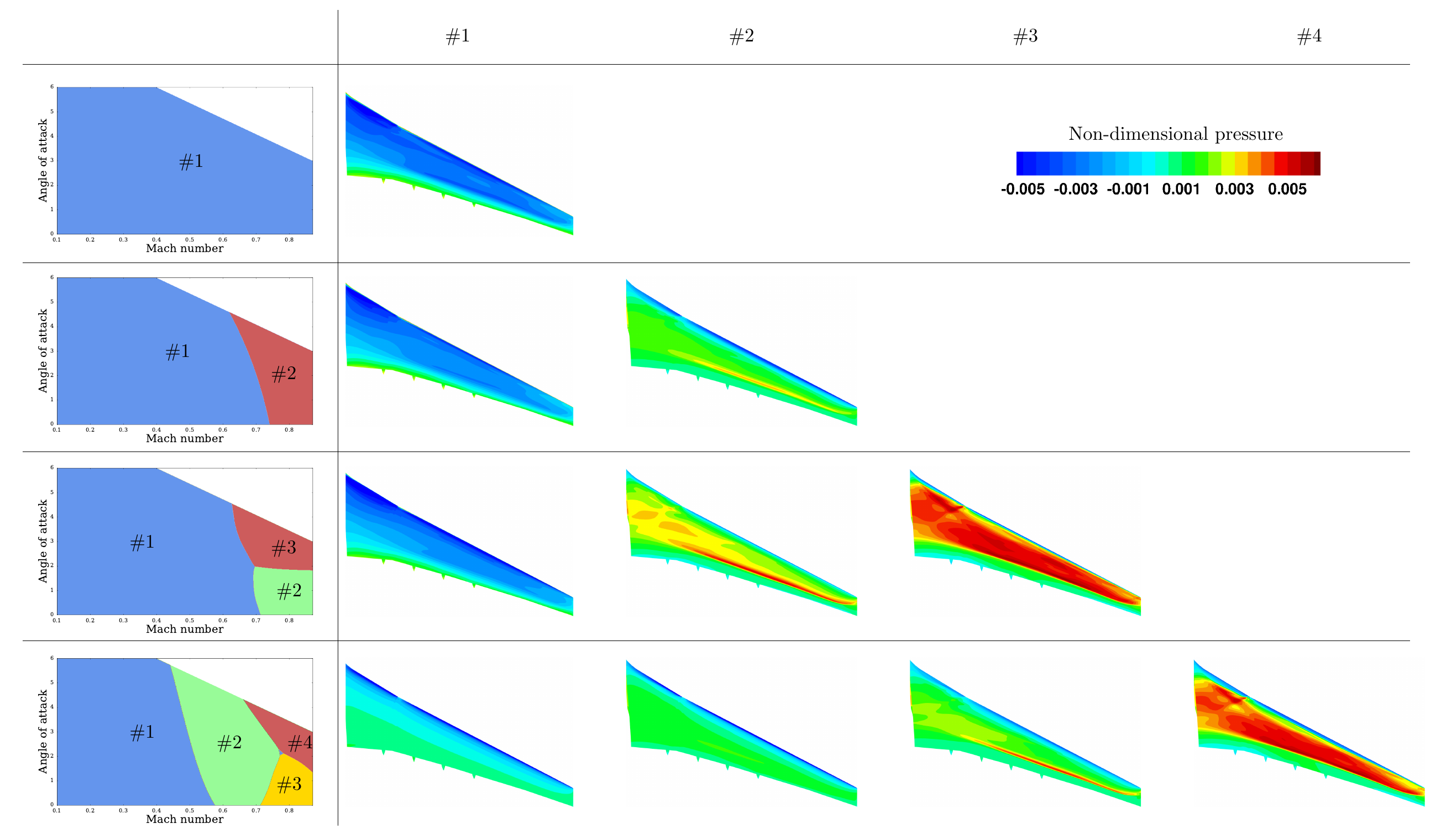}
\caption{First mode on each cluster at $70$ training samples.}
\label{fig:xrf1_first_mode}
\end{figure}
\end{landscape}

A more detailed analysis of the POD modes is now proposed.~\autoref{fig:xrf1_first_mode} represents the first POD mode on the suction side for different LDM models with $70$ training samples. The cluster denoted by $\#1$ is represented by the parameter space decomposition in the left column. Its associated mode depicts a typical shape that is found for models with $1$, $2$, and $3$ clusters: a negative amplitude is observed on the majority of the wing with the exception of a narrow band near the trailing edge generating a non negligible gradient, which can be interpreted as the recompression occurring for subsonic flows. The latter is not present for the LDM with $4$ clusters. Indeed, its cluster $\#1$ is smaller and does not contain snapshots with a Mach number above $0.6$, explaining the smoother evolution of the mode along the chord. Thus, it might be expected that models with $1$, $2$, and $3$ clusters are able to deal with subsonic and low shocks. Moreover, this first mode for global ROM can also explain why such a model has troubles to predict transonic shocks: this mode must be representative of the whole parameter space while similar modes for local models are only restricted to Mach number values between $0.1$ and $0.7$.

\pagebreak[4]
\global\pdfpageattr\expandafter{\the\pdfpageattr/Rotate 90}

\begin{landscape}
\begin{figure}[h!]
\includegraphics[width=1.45\textwidth]{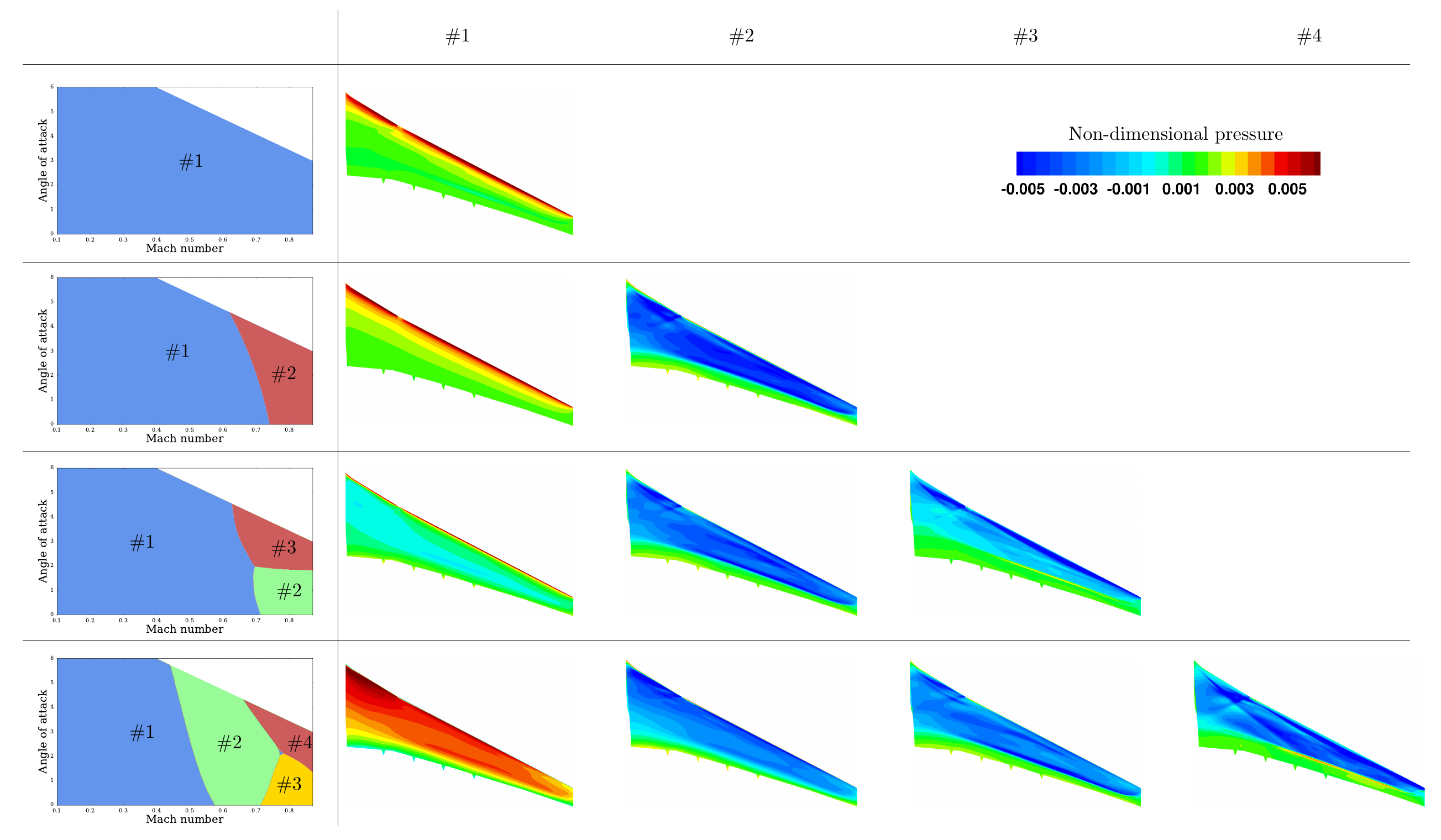}
\caption{Second mode on each cluster at $70$ training samples.}
\label{fig:xrf1_second_mode}
\end{figure}
\end{landscape}
\pagebreak[4]
\global\pdfpageattr\expandafter{\the\pdfpageattr/Rotate 0}

Models with $3$ and $4$ clusters show similarities in terms of parameter decomposition between respectively the clusters $\#2$ and $\#3$ and the clusters $\#3$ and $\#4$. Consequently, their POD modes share common shapes. In contrast with the cluster $\#1$, the red clusters are characterized by a strong discontinuity associated to positive values, while the clusters at a low angle of attack and high Mach number emphasize only narrow peak values located about the discontinuity observed for the other modes. This narrow peak can also be found in the second cluster of the model with $2$ clusters. So it is clear that various physical phenomena are contained in the POD modes. However they are not highlighted in the same way between the models. 

The second POD modes are presented in~\autoref{fig:xrf1_second_mode}. Only the clusters $\#1$ show different shapes, all the other clusters are very similar and look like the clusters $\#1$ of the first POD mode. Interestingly, the case with $4$ clusters differentiates itself from the other models with a peak positive value on nearly all the suction side counterbalancing its smooth first mode.

The POD analysis of the AS$28$G has provided a clear delimitation between subsonic and transonic flows with only $2$ clusters~\cite{Dupuis2018}. It seems that the application of the LDM to the XRF-$1$ configuration may require more clusters. Indeed, the clear separation between smooth and discontinuous modes is an expected result of the LDM and it encourages in particular for further study of models with more than $2$ clusters.

\subsection{Numerical results}
The accuracy of the LDM is presented in term of the normalized error for each snapshot $j$:
\begin{equation}
E_j = \frac{ \langle |\bm{f}(\bchi_{p_j}) - \widetilde{\bm{f}}(\bchi_{p_j})
|\rangle_{\Omega}} {f_{max} - f_{min}}, \ \forall j \in [1,m],
\label{eq:ei_def}
\end{equation}
where $f_{max}$ and $f_{min}$ refer respectively to the scalar maximum and
minimum values taken by the function $\bm{f}$. It corresponds to the absolute error between the exact value and the prediction, normalized by the range of variation, at snapshot level $j$. Using such error metrics aims at providing a statistical error analysis, the averaged normalized error $E_i$ is introduced. 

\begin{figure} [h!]
\centering
\begin{subfigure}{1\textwidth}
\centering
\includegraphics[width=1.0\linewidth, trim={0cm 0cm 0.5cm 0.35cm},clip]{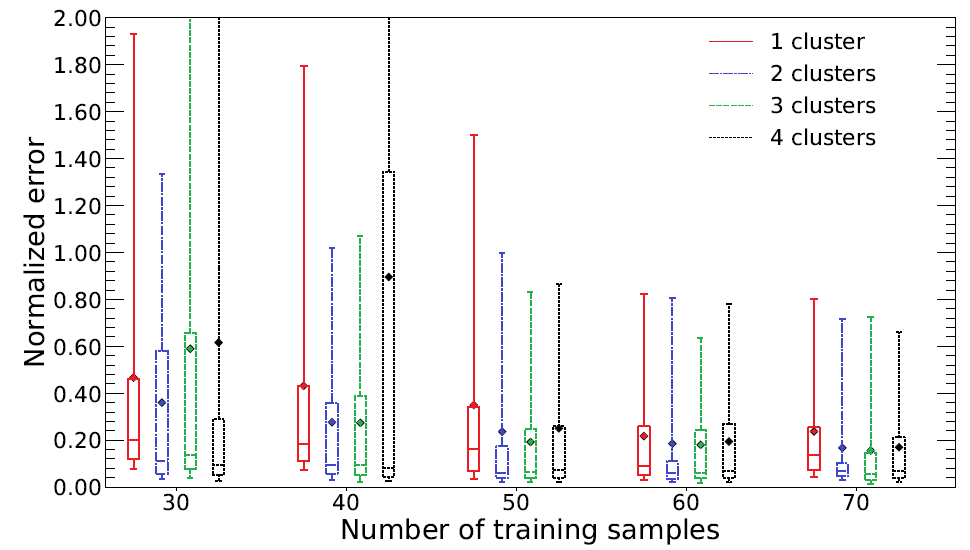}
\end{subfigure}
\caption{Box plots of the normalized error for different numbers of training samples.}
\label{fig:xrf1_results}
\end{figure}

~\autoref{fig:xrf1_results} depicts the evolution of the error with the number of training samples in a box plot formalism (see \autoref{sec:app_box_plot} for details). The error is computed using $200$ test samples. First of all, the main remark to be made regards the model with $2$ clusters. The latter outperforms the global model whatever the number of training samples. This observation is particularly true for $30$, $40$, and $50$ training samples where LDM reduces maximum, median, and mean error by a significant value.

Moreover, as predicted with the LOO error, the LDM with $3$ and $4$ clusters shows a poor accuracy with $30$ training samples (and $40$ training samples for the model with $4$ clusters). Beyond this threshold, they reach an acceptable performance level. Nevertheless, when the number of training samples increases, the global method reduces the gap with the LDM. For $60$ and $70$ training samples, maximum errors are similar and LDM reduces the mean error by about $0.04$. This latter may seem low because the global ROM shows good accuracy, but in relative value the mean error is reduced by $20\%$ even for $60$ and $70$. One can also note that the interquartile error is much lower for the LDM with $2$ clusters than the global ROM.

Two possible interpretations are proposed to explain the two different behaviors: (i) above $50$ training samples, a sufficient number is reached and can overcome limitations of the global model, as regression models are well trained and POD modes are representative, (ii) below $50$ training samples, an important number of clusters is incompatible with a low number of training samples, and (iii) the truly challenging points are too few in the test sets, since the parameter space is evenly filled. In particular, this last point highlights the need for a specific process taking into account the difference between the physical regimes. For instance, a specific resampling strategy could be used for the test sets instead of a uniform one. Such an approach could provide further details to understand the difference between the configuration with $2$, $3$, and $4$ clusters.

Nevertheless, it can be concluded that the XRF-1 has been widely enhanced by the LDM, in particular with $2$ clusters. This improvement is mitigated with $3$ and $4$ clusters, but the XRF-1 is a complex configuration and it will be interesting to continue the analysis beyond $70$ training samples.
\section{Conclusion and main findings}
\label{sec:conclusion}
An improved version of the LDM was successfully applied and assessed with simulations of turbulent flows around the industrial-like XRF-1 configuration. Seven different parameters were considered, including the engine conditions. Observations revealed a
consistent separation of the parameter space with continuous regions and clear identification of the subsonic and transonic regime. Moreover, a significant improvement of accuracy compared to the classical method was reached for almost all cases. These results were expected as the LDM was designed to distinguish between subsonic and transonic flows. In particular, the relevance of the resampling step was highlighted, as subdomains associated to high gradients and discontinuities were targeted.

Nevertheless, the LDM emerged as a sophisticated method that may require to tune several numerical parameters. Its application on the XRF-1 case was straightforward with impressive results with $2$ clusters. However, several tools must be deployed to ensure the quality of the result, in particular the analysis of the clustering (Silhouette analysis, local leave-one-out, etc.) and specific handling of the interface between the local models. These tools are particularly important for complex applications such as the XRF-1.

Finally, this aerodynamic test cases emphasized the complexity to validate surrogate models. For XRF-1, more than $150$ simulations were performed with a significant CPU cost. However, measuring the real accuracy of each model appeared very challenging, mainly due to: the size of the parameter space, the computational costs, and the vast amount of data, as each simulation represents thousands of scalars. Finding a relevant error metric represents a major challenge as the flow is very different within the parameter space. Deep analyses of the sampling, the POD modes, and the error distribution using a box plot formalism represent a first step but further investigations remain mandatory to provide a clear, robust, and reliable assessment of non-intrusive ROMs.
\appendix

\section{Appendix: box plot formalism}
\label{sec:app_box_plot}
\begin{figure}[ht]
\centering
\includegraphics[width=0.4\textwidth]{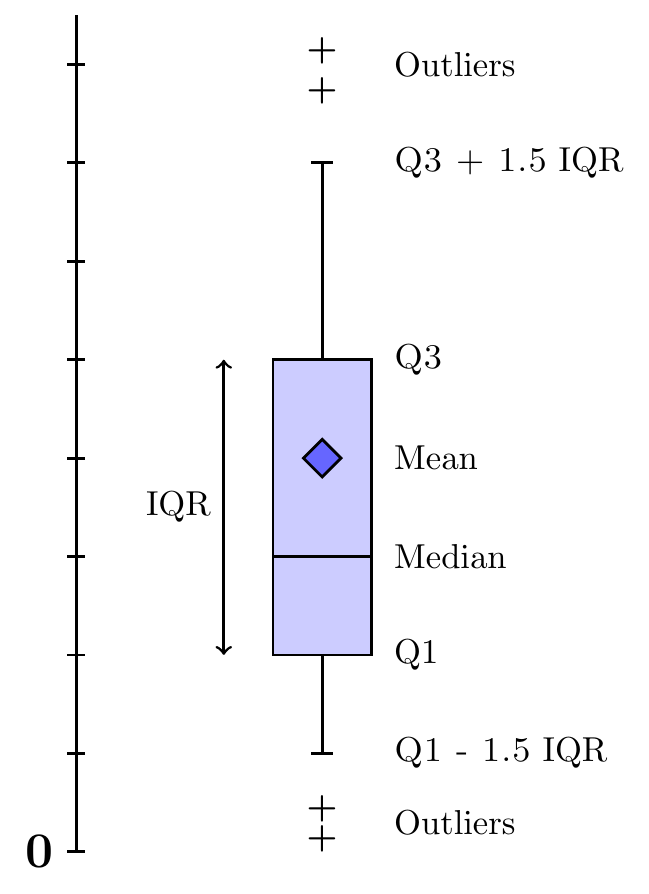}
\caption{Illustration the box plot representation.}
\label{fig:box_plot}
\end{figure}

An example of box plot is illustrated in~\autoref{fig:box_plot}. A
box plot groups the data through different quantiles: the bottom and the top of
the box represent respectively the value of the first and third quartiles Q1 and Q3,
whereas the horizontal line inside the box is the median (second quartile) and
the diamond the mean. The distance between the two quartiles Q1 and Q3 denotes the interquartile range of the data (IQR). The first vertical line indicates minimum value equal to Q1$- 1.5$ IQR and the last vertical line gives the extreme value associated with Q3$+1.5$ IQR. Finally outliers are plotted as crosses or dots.
Using box plot formalism to assess the ROM error allows to visualize in detail the distribution of the error for each snapshot. This approach provides more information than classical averaged values.

\section*{Acknowledgments} 
This work is part of the MDA-MDO project of the French Institute of Technology
IRT Saint Exupery. We wish to acknowledge the PIA framework (CGI, ANR) and the
project industrial members for their support, financial backing and/or own
knowledge: Airbus, Airbus Group Innovations, SOGETI High Tech, Altran Technologies, CERFACS.
The studies presented in this article is making use of the \textit{elsA} -ONERA software, whose the co-owners are Airbus, Safran, and ONERA. The authors would like to thank Airbus for providing the XRF-1 test case as a mechanism for demonstration of the approaches presented in this paper.
\bibliographystyle{aiaa}
\bibliography{references}

\end{document}